\documentclass[a4paper,12pt]{article}
\usepackage[utf8]{inputenc}
\usepackage{geometry}
\usepackage[english]{babel}
\usepackage[onehalfspacing]{setspace}
\usepackage[normalem]{ulem}
\usepackage{tikz}
\usetikzlibrary{calc,shapes,backgrounds,arrows,automata,shadows,positioning, petri}

\usepackage{geometry}
 \usepackage[pagewise]{lineno}
 
\definecolor{redorg}{RGB}{215,48,39}
\definecolor{orangeorg}{RGB}{253,174,97}
\definecolor{blueind}{RGB}{69,117,233}
\definecolor{cyanind}{RGB}{116,173,209}
\definecolor{greenind}{RGB}{171,217,233}
\usepackage{amsmath,amssymb,amsthm}
\usepackage{apxproof}
\usepackage{xcolor}
\usepackage{natbib}
\usepackage{ifthen}
\usepackage{authblk}
\usepackage{xr}
\usepackage{hyperref}
\newcommand{\inc}{A}
\newcommand{\pcl}{Z}
\newcommand{\scl}{W}
\newcommand{\btheta}{\boldsymbol{\theta}}

\newcommand{\param}{\btheta}
\newcommand{\nr}{n_{r}}
\newcommand{\nc}{n_{c}}
\newcommand{\qr}{Q_{r}}
\newcommand{\qc}{Q_{c}}
\newcommand{\pmix}{\pi}
\newcommand{\smix}{\rho}
\newcommand{\con}{\delta}

\newcommand{\rob}{R}

\newcommand{\Sbb}{\mathbb{S}}
\newcommand{\Ubb}{\mathbb{U}}
\newcommand{\Blbb}{\mathbb{B}}
\newcommand{\BbbU}{\mathbb{B}^{\uparrow}}
\newcommand{\BbbD}{\mathbb{B}^{\downarrow}}
\newcommand{\Dbb}{\mathbb{D}}

\newcommand{\DbbUo}{\mathbb{D}^{\uparrow}_{ord}}
\newcommand{\DbbDo}{\mathbb{D}^{\downarrow}_{ord}}
\newcommand{\DbbUa}{\mathbb{D}^{\uparrow}_{lin}}
\newcommand{\DbbDa}{\mathbb{D}^{\downarrow}_{lin}}

\newcommand{\robasm}[3]{R(#1, #2, #3)}

\newcommand{\robam}[3]{R_{#2}(#1, #3)}
\newcommand{\robm}[3]{R_{#1,#2}(#3)}
\newcommand{\robas}[2]{\overline{R}(#1, #2)}

\newcommand{\roba}[2]{\overline{R}_{#2}(#1)}
\newcommand{\robst}[2]{\overline{R}_{#1,#2}}
\newcommand{\lbm}[2]{#1, #2}
\newcommand{\pr}{\mathbb{P}}
\newcommand{\E}{\mathbb{E}}
\newcommand{\Symnr}{\mathfrak{S}_{\nr}}
\newcommand{\var}{\mathbb{V}}

\newcommand{\bn}{\boldsymbol{n}}

\newcommand{\bgamma}{\boldsymbol{\gamma}}

\newcommand{\OA}[2]{\sout{\textcolor{gray}{#1}}\textcolor{black}{#2}}

\newtheoremrep{proposition}{Proposition}
\newtheoremrep{corrolary}{Corrolary}
\newtheoremrep{definition}{Definition}
\newtheoremrep{properties}{Properties}
\newtheoremrep{cbnlem}{Combinatorial lemma}
\newtheoremrep{remark}{Remark}

\title{Impact of the mesoscale structure of a bipartite ecological interaction network on its robustness through a probabilistic modeling \thanks{Accepted for publication at Environmetrics}}
\author[1]{Saint-Clair Chabert-Liddell \thanks{Corresponding author, \texttt{saint-clair.chabert-liddell@inrae.fr}}}
\author[1]{Pierre Barbillon}
\author[1]{Sophie Donnet}
\affil[1]{Université Paris-Saclay, AgroParisTech, INRAE, UMR MIA-Paris, 75005, Paris, France}

\date{}

\begin{document}
\maketitle

\begin{abstract}    
    The robustness of an ecological network quantifies the resilience of the ecosystem it represents to species loss. It corresponds to the proportion of species that are disconnected from the rest of the network when extinctions occur sequentially. Classically, the robustness is calculated for a given network, from the simulation of a large number of extinction sequences.  The link between network structure and robustness remains an open question.  Setting a  joint probabilistic model on the network and the extinction sequences allows \OA{}{analysis of} this relation. 
    Bipartite stochastic block models have proven their ability to model 
bipartite networks e.g. plant-pollinator networks: species are divided into blocks and interaction probabilities are determined by the blocks of membership.  Analytical expressions of the expectation and variance of robustness are obtained under this model, for different distributions of primary extinction sequences. The impact of the network structure on the robustness is analyzed through a set of properties and numerical illustrations. The analysis of a collection of bipartite ecological networks allows us to compare the empirical approach to our probabilistic approach, and illustrates the relevance of the latter when it comes to computing the robustness of a partially observed or incompletely sampled network.

    
\end{abstract}

\section{Introduction}

In response to the rapid evolution of ecosystems due to climate change (habitat losses and species extinctions) on the one hand, and the increasing number of available data sets of ecological interaction networks on the other hand, the study of the robustness of ecological networks to species loss has become an active area of research in the ecological scientific community \citep[see][for a review]{landi2018complexity}. Given a primary extinction sequence, the influence of these extinctions on the other species  is studied  by monitoring the proportion of species that get disconnected from the rest of the network \citep{dunne2002network}.

Bipartite networks are used to represent interactions between two separated sets of species that do not interact within their own set. These interactions may be either mutualistic  when the species of both groups benefit from the interactions (such as pollination for plant-pollinator networks or seed-dispersal for plant-frugivore bird networks) or antagonistic when one group of species benefits from the interactions at the expense of the other group of species (e.g.  plant-herbivore or host-parasite networks).  
Robustness  is a numeric indicator quantifying the   impact of the disappearance of one set of (primary) species on the other (secondary species) under given extinction sequences \citep[see][for examples of primary extinction sequences]{memmott2004tolerance, curtsdotter2011robustness}. 
In a few words, the object of interest is the proportion of secondary species that remain connected to \OA{}{other species} after a given number of primary species have disappeared. Since this proportion depends on the order in which the species disappear, robustness is defined as the average of these proportions over a large number of primary extinction sequences. 
\footnote{Note that the robustness of ecological networks bears a different meaning than the robustness traditionally used in the information or epidemic networks literature. In these fields, the goal is to see how a network stays connected when some nodes are disabled (or on the opposite how epidemics are still able to spread when nodes become immune), the task is tackled by analyzing the size of the different connected components and in particular the existence of a giant connected component \citep[see][chapter 15 and reference therein]{newman2018networks}. In ecology,   the key issue is to observe   how species get isolated from the rest of the ecosystem.
This approach is  related to the concept of isolated nodes in a graph \citep{erdHos1960evolution}.}  

\OA{}{
The definition of ecological robustness used in this paper is not the only one, in particular the robustness of ecological networks is also studied through dynamic approaches. Among recent developments in the dynamic approach, \citet{song2018guideline} are interested in the feasibility domain of ecological communities, i.e. the set of environmental conditions under which all species have positive abundances.  \cite{barabas2014sensitivity} estimate extinction risk to small environmental variation through the use of sensitivity analysis  while others study the capacity of coexistence of species through the so-called invasion criterion  \citep[see][for a review]{grainger2019invasion}.
}

 Obviously, \OA{}{the robustness defined above} is strongly related to the size of the network considered (i.e. the number of species involved or \emph{species richness}) and to its density (i.e. the number of interactions observed compared to the total number of possible interactions or \emph{connectance}). To go further, the question arises to what extent the topological properties of a network also influence its robustness. 
Indeed, observed ecological networks present different topological structures depending on the type of interactions: for example, mutualistic networks are known to have a strong nested structure, while this is not necessarily the case for antagonistic networks \citep{fortuna2010nestedness, bascompte2003nested}. 
A key question is to identify the relationship between the ability of a network to withstand species extinctions and its characteristics such as its size and its mesoscale structure. 

To that purpose, we propose to assume a parametric probabilistic model for the bipartite network. This model will embed the topological properties of the network in a few number of parameters.
Then we propose to study the behavior of the robustness in this framework and in particular the variations of the robustness with respect to the model parameters.
More precisely, we focus on the expectation and the variance of the robustness under a network probabilistic model.  For some particular probabilistic models and some particular extinction sequences, the relation between the parameters and the expected robustness can be provided in a closed-form, making a fine study of the relation possible.

Furthermore, relying on a probabilistic model which can be adjusted to account for an observation process may enable \OA{}{for correction of the sampling} effect on the robustness.  Indeed, even though ecological networks are often considered to describe all the possible interaction\OA{}{s} between species, the sampling may be incomplete
\citep{bluthgen2008interaction} which may bias the computed network statistics \citep{rivera2012effects, de2020biased} such as the robustness.

The stochastic block model \citep[SBM][]{nowicki2001estimation} and its extensions for bipartite networks, such as the biSBM \citep[also referred \OA{}{to} as Latent Block Model]{govaert2003clustering} or its degree corrected counterpart \citep[DCbiSBM][]{larremore2014efficiently} have gained a lot of attention in the statistical and network science research fields during the last decade.  While the Erd\H{o}s-R{\'e}nyi model \citep{erdHos1960evolution} assumes that any pair of species has the same probability \OA{}{of} interaction, SBMs introduce heterogeneity in the connection behavior. Specifically, in bipartite SBMs, the two sets of nodes are divided into clusters/blocks/groups and the probability that two nodes are connected depends on which blocks the nodes belong to. Depending on the number of blocks, their size, and interaction probabilities (inter- and intra-block), SBMs encompass a wide variety of topologies (such as assortative community or core-periphery) and encode them in a small number of parameters.  

Since \cite{allesina2009food} has advocated for the use of groups in ecological networks, SBMs (sometimes referred to as group models) have gained in popularity. Some variants adapted to multilayer ecological networks have been proposed\OA{}{for:} multiplex networks \citep{kefi2016structured}, multipartite networks \citep{bar2020block} or temporal networks \citep{matias2017statistical}. Besides, they have been used  to answer specific ecological questions. To name a few, \citet{ michalska2018understanding} explore the structural role of parasite species in food webs, while 
\citet{miele2020core} use biSBM to assess the core-periphery structure of plant-pollinator networks.
Furthermore, SBMs provide an ecological interpretation in terms of  functional groups in the ecosystem: species in \OA{}{the} same block interact in a similar way, which means that the exchangeability of species in a block model is related to the concept of ecological equivalence \citep{sander2015can}.

To our knowledge,  the behavior of the robustness has never been studied  in the SBM framework.
Some grouping algorithms are used to derive extinction sequences.
\cite{cai2016robustness} optimize an objective function to determine communities in the network and then generates primary extinction based on these communities.
More generally, some models are used to generate   primary extinction sequences, or to model how species of the other functional group react to those extinctions \citep[such as rewiring or cascading, see][for examples]{bane2018effects, vizentin2020including}.
In an approach more similar to ours, \citet{burgos2007nestedness} stud\OA{}{ied} the relationship between nestedness and robustness of mutualistic networks by using the self-organizing network model \citep{medan2007analysis} to generate nested networks and deriv\OA{}{ed} analytical expression of the robustness under this model.

The robustness given as a decreasing function and a robustness statistic classically used in ecology are \OA{}{described} in Section \ref{sec:robustness}. 
Section \ref{sec:model} is dedicated to the introduction of the biSBM, the DCbiSBM and related models for sequential species extinctions.  Section \ref{sec:moment} supplies the expression of the expectation and the variance of the robustness under a biSBM for different distributions of primary extinction sequences. The  analytical properties  of the expected robustness  are given in Section \ref{sec:properties} together with more general studies to illustrate the  impact of the network structure on the robustness.
In Section \ref{sec:appli}, we apply our approach on a dataset composed of both mutualistic and antagonistic bipartite networks and compare our results to the classical approach.  Finally, on the same dataset, we show how our approach allows us to calculate the expected robustness of partially observed ecological networks.

\section{Robustness of bipartite ecological networks}\label{sec:robustness}

\OA{}{R}obustness aims at measuring the tolerance of a network to species extinctions by quantifying the proportion of remaining species along a species extinction sequence.
Formally, let $\inc \in \{0,1\}^{\nr \times \nc}$ be the $\nr \times \nc$ incidence matrix of a bipartite network representing ecological interactions between two groups of species of respective sizes $\nr$ and $\nc$ (such as plant-pollinators or hosts-parasites). Then:   

\begin{eqnarray*}
   \inc_{ij} = \begin{cases}
                 1  & \text{ if row species $i$ interacts with column species $j$,} \\
                 0  & \text{otherwise}.
               \end{cases}
\end{eqnarray*}

Let $s$ be an   extinction sequence on the row species: $s \in  \Symnr$ where $\Symnr$ is the symmetric group. The extinctions in row  (whose order is given by $s$) are the \emph{primary extinctions}. These row extinctions lead to secondary extinctions among the column species if these column species remain isolated after the disappearance of row species. More precisely, a column species $j$  is said to be be extinct after $m$ row primary extinctions if these $m$  primary extinctions caused species $j$ to lose all its connections, or equivalently if after these $m$ primary row extinctions, $j$ has no connections left with the remaining rows, which is equivalent to $ \sum_{i=m+1}^{\nr} \inc_{s(i)j}=0$.  

For a given sequence $s$ and a given number of primary extinctions $m$,  $\robasm{\inc}{s}{m}$  is the proportion of remaining column species: 
\begin{equation}\label{eq:Rasm}
\robasm{\inc}{s}{m}  = 1 - \frac{1}{\nc}\sum_{j = 1}^{\nc} \mathbf{1}_{\{ \sum_{i=m+1}^{\nr} A_{s(i)j} = 0 \}}.
\end{equation}
Note that for all $m\in\{0,\ldots,n_r\}$, $0\le \robasm{\inc}{s}{m}\le 1$ with
$\robasm{\inc}{s}{n_r}=0$.
The \emph{robustness function} is defined as  the expectation of 
$\robasm{\inc}{S}{m}$ against the primary extinction sequences: 
\begin{equation}\label{eq:rob_function2}
 {m} \mapsto \robam{\inc}{\Sbb}{m}= \E_{S}\left[\robasm{\inc}{S}{m}\right] \quad \mbox{ with } \quad S  \sim  \Sbb  \,.
\end{equation}
where $\Sbb$ is a probability distribution on  $\Symnr$. 
Equation \eqref{eq:rob_function2} is a weighted summation over the $\nr!$ possible permutations of $\Symnr$, which may render the computation intractable. In practice,  it is  generally approximated by a Monte Carlo integration: 
\begin{equation}\label{eq:rob_mc}
 \widehat{\rob}_{\Sbb}\left(\inc,m\right) = \frac{1}{B}\sum_{b=1}^B \robasm{\inc}{S^{(b)}}{m}\quad \mbox{ where } \quad S^{(b)} \sim_{i.i.d}  \Sbb,  \quad  b=1,\dots, B.  
\end{equation}
The \emph{robustness statistic} of  a network $\inc$ and a primary extinction sequences distribution $\Sbb$ is defined as:
\begin{equation}\label{eq:rob_stat2}
        \roba{\inc}{\Sbb} = 
        \frac{1}{\nr}\sum_{m=0}^{\nr} \robam{\inc}{\Sbb}{m},
\end{equation}
which corresponds to the area under the curve (AUC)  of the robustness function where the x-axis has been re-normalized to match with the proportion of removed row species. $\widehat{\overline{R}}_{\Sbb}(\inc)$ is the Monte Carlo version of $\roba{\inc}{\Sbb}$. 

Other statistics exist in the literature, such as the median of the robustness function \citep{dunne2002network}, i.e. the proportion of primary extinctions needed to provoke a secondary extinction on half of the species. However, the statistic  \eqref{eq:rob_stat2} is widely used in \OA{}{ecology} and is mathematically convenient for our purpose. 

\paragraph{About the extinction sequence distribution $\Sbb$}
Several choices for $\Sbb  $ corresponding to various ecological scenari\OA{}{os} are commonly considered in the literature.  The first  choice is  the  uniform distribution over $\Symnr$ ($\Sbb = \mathbb{U}_{_{\Symnr}}$), assuming that the species disappear  without specific order. For the sake of simplicity, we use $\mathbb{U}$  instead of $\mathbb{U}_{\Symnr}$.  

Distributions $\Sbb$  depending on  $A$  are suggested in the literature \citep[see][for examples]{curtsdotter2011robustness}.  
Among these approaches,  we will focus on  sequences depending on the row degree sequences. For any row species  $i = 1,\dots, \nr$, let $D_i = \sum_{j=1}^{\nc} A_{ij}$ be  its  degree, i.e. the number of edges involving $i$. 
On the one hand, the worst-case scenario in terms of ecological extinctions assumes that the row  species with the highest degrees (most connected row species) disappear first. In this case, the species in rows are ordered in decreasing degrees and the primary extinction sequences follow this order;  row species of equal degrees disappear in a uniformly distributed order. 
On the other hand, the generation of primary extinction sequences that first eliminate species of lower degree mimics a more favorable ecological scenario. 
In these two cases, the support of $\Sbb$  is of cardinal $\prod_k \#\{i:D_i = k\}!$.  Depending on the sequence $(D_i)_{i=1,\dots,\nr}$, the corresponding $\robam{\inc}{\Sbb}{m}$  may become tractable or not. If not, a Monte Carlo approximation is used.   

Instead of considering a strict monotone ordering of the row degrees, one might relax the constraint  and set the probability for a row species to disappear proportional to a function of its degree, as seen in \citet{liu2019node}. This would correspond to sampling without replacement a sequence $s$ where the weights of each species $i \in \{1, \dots, \nr \}$ are, for instance given by: 
  \begin{equation}\label{eq:weight_degree}
    w_i \propto  D_i^{\alpha}\,.
  \end{equation}
 $\alpha = 0$  coincides to the uniform distribution $\mathbb{U}$ while, if $\alpha = 1$, the primary extinction sequence depends linearly on the degrees.  
 The increasing order is obtained by reversing this sequence of primary extinction.
 
\begin{remark}
The  $\robam{\inc}{\Sbb}{m}$'s computed for the  first  two $\Sbb$'s described above are the most widely used. They are implemented in the \texttt{R} package \texttt{bipartite} \citep{dormann2008introducing} available on \texttt{cran}.  
 
\end{remark}

The robustness function \eqref{eq:rob_function2} and corresponding statistic \eqref{eq:rob_stat2}  are then computed  from a unique observed network $A$ and a probability distribution  $\mathbb{S}$ conditional to $A$ (through the row degrees): 
\begin{equation}\label{eq:rob_function}
 \robam{\inc}{\Sbb}{m}= \E_{S | A}\left[\robasm{\inc}{S}{m}\right] \quad \mbox{ with } \quad S | A  \sim  \Sbb  \,.
\end{equation}
This robustness computed for a given $A$ will be \OA{}{referred to} as the \emph{empirical robustness}. 

 In order to understand the variability of the robustness with respect to the network structure  we  set a probabilistic model on the network $A \sim \mathbb{A}_\theta$ where the parameters $\theta$ embeds the network structure in a small number of parameters and then study the random variable $\robam{\inc}{\Sbb}{m}$ for various distributions on 
$(S,A)$.  
The following section is dedicated to the description of flexible probabilistic models on $A$ and  adapted joint distributions on $A$ and $S$.

\section{Bipartite Stochastic Block Model and related sequential extinctions }\label{sec:model}

\subsection{Probabilistic model on bipartite ecological networks}

\paragraph{Bipartite stochastic block models}

The bipartite SBM, a.k.a. the Latent Block Model \citep{govaert2003clustering}, is a mixture model on the edges   adapted to bipartite networks. It relies on a clustering of nodes in rows and a clustering of nodes in columns. The nodes which belong to the same cluster (or equivalently block) are assumed to share the same connectivity profile in the network and the probability of interaction between two nodes depends on the blocks they belong to. More precisely, each of the $\nr$ row species is attributed to a block $k \in \{1,\dots, \qr\}$ independently from the other species. Let $\pcl_i$ be such that $\pcl_i = k$ if row species $i$ belongs to block $k$. The $\pcl_i$'s are assumed to be independent and identically distributed (i.i.d.) and:
\begin{equation}\label{eq:mix_row}
    \pr (\pcl_{i} = k) = \pmix_k \quad i \in \{1, \dots, \nr\},
\end{equation}
with $\sum_{k=1}^{Q_r} \pi_k =1$. For $j=1,\dots, \nc$, let $\scl_j$ be such that $\scl_j = q$ if column species $j$ belongs to block $q \in \{1, \dots, \qc\}$.   The $\scl_j$'s are assumed to be i.i.d. and:
\begin{equation}\label{eq:mix_col}
    \pr (\scl_{j} = q) = \rho_q \quad j \in \{1, \dots, \nc\},
\end{equation}
with $\sum_{q=1}^{Q_c} \rho_q =1$. 
Then, conditionally to their respective  latent blocks, the interactions between two species are distributed independently as:  
\begin{equation}\label{eq:Y|Z}
    \inc_{ij} | \{\pcl_{i}=k, \scl_{j}=q\} \sim \mathcal{B}(\con_{kq}), 
\end{equation}
where $\mathcal{B}$ is the Bernoulli distribution and $\con_{kq} \in (0,1)$ for $(k,q)\in\{1,\ldots,Q_r\}\times\{1,\ldots,Q_c\}$. 
The parameter $\btheta = \{\rho_q, \pi_k,\con_{kq}\}_{k=1,\dotsc, Q_r, q= 1,\dotsc,Q_c}$  then encodes the topology of the network. 
Let $\boldsymbol{n}=(\nr ,\nc)$ be the numbers of species (richness) in rows and in columns of the network: $  biSBM(\btheta,\bn)$ denotes the distribution on the networks defined by equations \eqref{eq:mix_row}, \eqref{eq:mix_col} and \eqref{eq:Y|Z} for a given value of $\btheta$.
 Note that $d = \sum_{k,q} \pi_k \delta_{kq} \rho_q$ is the expected connection probability for any pair of species $(i,j)$.  $d$ will be referred to as the expected density. .

\begin{remark} A special case  among the biSBM distributions is the one where $\delta_{kq} = d$ for any $k,q$, or equivalently --in terms of models on networks-- where $\qr = 1$ and $\qc = 1$. In that case, all connections occur  independently with the same probability $d$.
This biSBM is the bipartite Erd\H{o}s-Rényi network. 
\end{remark}

\paragraph{Degree corrected stochastic block models}

The degree corrected stochastic block model \citep{karrer2011stochastic}
is an extension of the SBM which accounts for the heterogeneity in the degree distribution. This heterogeneity does not only depend on the blocks but also on some parameters related to the nodes themselves.
The extension to bipartite network is done in \citet{larremore2014efficiently}.
In these models, the distribution on the dyads is a Poisson distribution even if the observed network consists of binary edges. This is a reasonable approximation in the case where the network is large and sparse. 
However, in ecological interaction networks, this assumption is hardly met. That is why we use a true binary definition of the degree corrected biSBM that we denote by DCbiSBM in the following.
On top of the blocks for the nodes in rows and columns given by variables $Z$s and $W$s as defined in Equations \eqref{eq:mix_row} and \eqref{eq:mix_col}, two vector parameters $\gamma_r\in\mathbb{R}^{n_r}$ and $\gamma_c\in\mathbb{R}^{n_c}$ are associated to the nodes and control their degree. \OA{}{ The larger $\gamma_{r,i}$ (resp. $\gamma_{c,j}$) the higher the probability of connection involving species $i$ (resp. $j$), compared to other species within the same block.}
Then, Equation \eqref{eq:Y|Z}
is replaced with
\begin{equation} \label{eq:Y|ZDC}
 \inc_{ij} | \{\pcl_{i}=k, \scl_{j}=q\} \sim \mathcal{B}(1/(1+\exp(-(\con_{kq}+\gamma_{r,i}+\gamma_{c,j})))\,,
\end{equation}
with $\gamma_{r,1}=\gamma_{c,1}=1$ for identifiability issues.

We denote by $DCbiSBM(\param,\boldsymbol{\gamma},\bn)$ where $\boldsymbol{\gamma}=(\gamma_r,\gamma_c)$, this probabilistic
model with a given set of parameters.
If there is only one block, i.e. $Q_r=Q_c=1$, this model only
accounts  for the heterogeneity of degrees and so
 corresponds to a bipartite version of the expected degree distribution (EDD) model \citep{chung2002connected}.

\subsection{Extinction sequence distributions adapted to bipartite Block Models}\label{sec:3_2}

As described in Section \ref{sec:robustness}, classically, the extinction sequences   are either distributed uniformly on  $\Symnr$ or conditionally to $A$. Positing a probabilistic distribution on 
$\inc$ leads to a joint distribution on  $\inc$ and $S$.
In the case of a uniform distribution on $S$:  $S \sim  \Ubb$, the joint distribution on $(\inc,S)$ consists of the product of the distributions for $\inc$ and $S$.
We will then denote by $\mathcal{L}_{\btheta,\bn,\Ubb}$ and  $\mathcal{L}_{\btheta,\boldsymbol{\gamma},\bn,\Ubb}$ the corresponding joint distributions when $\inc \sim biSBM(\btheta,\bn)$ or $\inc \sim DCbiSBM(\btheta,\bgamma,\bn)$ respectively.

As described in Section \ref{sec:robustness}, the extinction sequences may depend on $\inc$ through its degree distributions.
When $\inc$ follows a probabilistic distribution, such an extinction sequence is defined conditionally on the realization of $\inc$. We propose \OA{}{another} dependence between $\inc$ and $S$ through the row clustering variables $Z$s.
More precisely, we consider that species of row block $1$ disappear first, then  block $2$, etc. Formally, let $\mathbb{B}$ be the uniform probability on the row species  following a given ordering of the blocks, then $S|Z \sim \mathbb{B}$ if: 
  \begin{equation}\label{eq:Block extinc}
     \pr(S = s | \pcl) = \frac{\mathbf{1}_{\pcl_{s(1)} \leq \dots \leq \pcl_{s(\nr)}}}{\# \{s : \pcl_{s(1)} \leq \dots \leq \pcl_{s(\nr)}\}}.
  \end{equation}
The  support of $\mathbb{B}$ is restricted to  $\prod_{k=1}^{\qr}n_k!$ elements, where $n_k$ is the cardinal of block $k$. 
Equations  \eqref{eq:mix_row}, \eqref{eq:mix_col}, \eqref{eq:Y|Z} and \eqref{eq:Block extinc} define a joint probability distribution on $(\inc,S)$ such that   $\inc$ and $S$ are independent conditionally to $Z$. We denote $\mathcal{L}_{\btheta,\bn,\mathbb{B}}$  this joint distribution on $(\inc,S)$ when $\inc \sim biSBM(\btheta,\bn)$. 
 Under this joint distribution, if $\btheta$ is such that $\delta_{k+} = \sum_{q = 1}^{\qc} \rho_q \delta_{kq}$ is a decreasing (resp. increasing) sequence, then $\Blbb$ will generate sequences with an expected decreasing (resp. increasing) sequence of degrees, i.e. with decreasing (resp. decreasing) connectivity. In the following, these extinction sequences will be referred to as block-decreasing (resp. block-increasing).
 The blocks may also be used to generate extinctions by ecological traits that correspond to the blocks.

If $\inc \sim DCbiSBM(\btheta,\bgamma,\bn)$, 
the extinction sequences are not related  to expected degrees  anymore but it may still make sense to generate extinction sequences linked to ecological traits.

\section{Moments of the robustness statistic}\label{sec:moment}

Studying the distribution of $\robasm{\inc}{S}{m}$ or $\robas{\inc}{S} = \frac{1}{n_r}\sum_{m=1}^M \robasm{\inc}{S}{m}$ under  the joint distribution of $A$ and $S$ could be done by simulation, using the fact that the biSBM and DCbiSBM are  generative models. However, this comes at a computational cost. 
In this section, we prove that, when $(A,S)\sim\mathcal{L}_{\btheta,\bn,\Ubb}$ or $(A,S)\sim\mathcal{L}_{\btheta,\bn,\mathbb{B}}$ the first moments of the  robustness are tractable in a closed-form. The proof
relies on the exchangeability of the nodes under a biSBM and on the fact that the considered extinction sequences are adapted to this exchangeability.
When  $\inc$ follows a DCbiSBM, the nodes are no longer exchangeable because of the parameters $\bgamma$ associated with each node.
Although we are able to obtain a closed-form expression, it is not tractable. Therefore, we will rely on a Monte Carlo approximation for summing on the extinction sequences.\\

We  now exhibit explicit expressions of
$\robm{\lbm{\param}{\bn}}{\Sbb}{m} = \E_\inc[\robam{\inc}{\Sbb}{m}]$ when $(\inc,S) \sim \mathcal{L}_{\btheta,\bn,\Ubb} \mbox { or } \mathcal{L}_{\btheta,\bn,\Blbb}$ and of 
$ \mathbb{V}_{\inc}[\robam{\inc}{\Ubb}{m})]$ when $\inc \sim biSBM(\btheta,\bn)$.

\subsection{Expectation} 

\begin{proposition}
\label{prop:rob_unif}
Let $(\inc, S) \sim \mathcal{L}_{\btheta,\bn,\Ubb}$ and set $\con_{+q} = \sum_{k=1}^{\qr}\pmix_{k}\con_{kq}$. Then,
\begin{itemize}
 \item $\forall m=1,\dotsc,\nr$:
\begin{eqnarray}\label{eq:rob_fun_unif}
        \robm{\lbm{\param}{\bn}}{\mathbb{U}}{m}&=&  1 - \sum_{q=1}^{\qc} \smix_q (1-\con_{+q})^{\nr-m} , 
\end{eqnarray} 
\item 

Consequently, the  robustness statistic is:
  \begin{eqnarray}\label{eq:auc_rob_unif}
  \robst{\lbm{\param}{\bn}}{\mathbb{U}}& = &\frac{1}{\nr}\sum_{m=0}^{\nr}    \robm{\lbm{\param}{\bn}}{\mathbb{U}}{m} =  1-\frac{1}{\nr}\sum_{q=1}^{\qc} \smix_{q} \frac{(1 - \con_{+q}) - (1-\con_{+q})^{\nr+1}}{\con_{+q}}. 
\end{eqnarray}
\end{itemize}
\end{proposition}
\begin{proof} 
 Since $(\inc,S) \sim \mathcal{L}_{\btheta,\bn,\Ubb}$,
\begin{eqnarray*}
\robm{\inc}{S}{m} & = & \E_{\inc}[\E_{S}[\,\robasm{\inc}{S}{m}\,] ]= \E_{\inc}[\,\robam{\inc}{\mathbb{U}}{m}\,] = \frac{1}{\nr!} \sum_{s \in \Symnr} \left(1 - \frac{1}{\nc}\sum_{j=1}^{\nc}\E_{\inc}\left[\mathbf{1}_{\sum_{i=m+1}^{\nr} \inc_{s(i)j}=0}\right] \right).
\end{eqnarray*}
Using the property of exchangeability  of the biSBM, we have that, for any permutation $s$, 
$\E_{\inc}\left[\mathbf{1}_{\sum_{i=m+1}^{\nr} \inc_{s(i)j}=0}\right] = \E_{\inc}\left[\mathbf{1}_{\sum_{i=m+1}^{\nr} \inc_{ij}=0}\right]$. 
Moreover, introducing the latent variables $\pcl$ and $\scl$ we have: 
\begin{align*}
 \E_{\inc}\left[\mathbf{1}_{\sum_{i=m+1}^{\nr} \inc_{ij}=0}\right] 
 &= \sum_{k_{m+1:\nr}\in\{1,\dotsc,\qr\}^{\nr-m}} \sum_{q=1}^{\qc}\pr_{}\left[\sum_{i=m+1}^{\nr} \inc_{ij}=0 | \pcl_{m+1:\nr} = k_{m+1:\nr}, \scl_{j} = q\right]\\
 & \quad  \quad  \quad \quad\quad \times \pr_{}\left(\pcl_{m+1:\nr} = k_{m+1:\nr}, \scl_{j} = q\right)\\
 &= \sum_{k_{m+1:\nr}\in\{1,\dotsc,\qr\}^{\nr-m}}\sum_{q=1}^{\qc}\prod_{i=m+1}^{n_r} (1-\con_{k_iq}) \left(\prod_{i=m+1}^{n_r} \pmix_{k_i}\right)\smix_q \\
 &= \sum_{q=1}^{\qc} \smix_q   \sum_{k_{m+1:\nr}\in\{1,\dotsc,\qr\}^{\nr-m}}\prod_{i=m+1}^{n_r} (1-\con_{k_i q}) \pmix_{k_i}=\sum_{q=1}^{\qc}  \smix_q \left( \sum_{k=1}^{\qr} \pmix_{k}(1 - \con_{kq})\right)^{\nr-m} \,.
\end{align*}
As a consequence, 
\begin{eqnarray*}
        \robm{\lbm{\param}{\bn}}{\mathbb{U}}{m} &=& 1- \sum_{q=1}^{\qc} \smix_q \left(\sum_{k=1}^{\qr} \pmix_{k}(1 - \con_{kq})\right)^{\nr-m} = 1 - \sum_{q=1}^{\qc} \smix_q (1-\con_{+q})^{\nr-m}\nonumber
\end{eqnarray*}
where $\con_{+q} = \sum_{k=1}^{\qr}\pmix_{k}\con_{kq}$. 
Then, averaging over $m$ leads to:
\begin{eqnarray*}
  \robst{\lbm{\param}{\bn}}{\mathbb{U}}& = &\frac{1}{\nr}\sum_{m=0}^{\nr}   \robm{\lbm{\param}{\bn}}{\mathbb{U}}{m} = 1-\frac{1}{\nr}\sum_{q=1}^{\qc} \smix_{q} \frac{(1 - \con_{+q}) - (1-\con_{+q})^{\nr+1}}{\con_{+q}}.
\end{eqnarray*}
\end{proof}

Note that, if the network has no specific structure ($\con_{kq} = d$ or $\qr = \qc = 1$) then   
$
\robm{\lbm{\btheta}{\bn}}{\Ubb}{m}  = 1- (1 - d)^{\nr-m}.    
$.

\begin{proposition}\label{prop:rob_block}
Let $(\inc,S) \sim \mathcal{L}_{\btheta,\bn,\Blbb}$, then
\begin{equation}\label{eq:robm n theta B}
  \robm{\lbm{\param}{\bn}}{\Blbb}{m} = 
    1 - \sum_{q=1}^{\qc} \smix_q \sum_{n_1 + \dots + n_{\qr} = \nr} \frac{\nr!}{n_1! \dots n_{\qr}!}  
\prod_{k=1}^{\qr}\pmix_{k}^{n_k} \left( 1-\con_{kq}\right)^{\min^{+}(n_k , \underset{l \leq k}{\sum}n_l - m)},
\end{equation}
where $\min^{+}$ is the positive part of the minimum function:  $\min^{+}(x, y) = \max(0, \min(x,y))$. 
\end{proposition}
The proof of Proposition \ref{prop:rob_block} is provided in the supplementary material \ref{S:proof:block}. 
The robustness statistic $ \robst{\lbm{\param}{\bn}}{\Blbb}$ is the mean of the  $\robm{\lbm{\param}{\bn}}{\Blbb}{m}$'s: no simplified expression has been obtained. 
Note that, if $\qr = 1$ or if $\con_{kq} = d$, then $\robm{\lbm{\param}{\bn}}{\Blbb}{m} = \robm{\lbm{\param}{\bn}}{\mathbb{U}}{m}$.

In Equation \eqref{eq:robm n theta B},  the summation over the partitions of the $\nr$ row species into $\qr$ blocks  may be burdensome  if $\nr$ or $\qr$ are large. In such cases, a Monte Carlo approximation may be used.  
Among the ecological networks we consider in Section \ref{sec:appli}, only a few of them require this approximation.

\subsection{Variance} 

We now aim at computing the variance of the robustness $ \mathbb{V}_{\inc}[\robam{\inc}{\Ubb}{m}]$, when $\inc \sim biSBM(\btheta,\bn)$, 
thus quantifying the variability of the robustness for a sample of networks sharing the same block models parameters (i.e. the same mesoscale patterns of connectivity).

\begin{proposition}\label{prop:variance}
 Let $\inc \sim biSBM(\btheta,\bn)$,  $\eta_q = 1-\con_{+q}$ and $\eta_{qq'} = \sum_{k=1}^{\qr} \pmix_k(1-\con_{kq})(1 - \con_{kq'})$.
 
 \begin{enumerate}
   \item Then:
  \begin{eqnarray*}
    \var_{\inc}[\robam{\inc}{\Ubb}{m}] &=&\frac{1}{\nc}   \sum_{l = m}^{\min(2m, \nr)}\frac{\binom{m}{l-m}\binom{\nr-m}{l-m}}{\binom{\nr}{m}}\sum_{q=1}^{\qr}\smix_{q}\eta_q^l - (\sum_{q=1}^{\qr}\rho_{q}\eta_q^m)^2 \\ 
        && + \frac{(\nc-1)}{\nc}  \sum_{l = m}^{\max(2m, \nr)}\frac{\binom{m}{l-m}\binom{\nr-m}{l-m}}{\binom{\nr}{m}}\sum_{q, q'=1}^{\qr}\smix_{q}\smix_{q'}(\eta_{q}\eta_{q'})^{l-m}\eta_{qq'}^{2m-l}\,.
\end{eqnarray*}
   \item The variance of the robustness statistic due to the network variability under a given biSBM is:
\begin{align*}
    \var_{\inc}&[\roba{\inc}{\Ubb}] = \frac{1}{\nr^2\nc^2} \nc \sum_{m, m'=0}^{\nr} \sum_{l = \max(m,m')}^{\min(m+m',\nr)}\frac{\binom{m}{l-m'}\binom{\nr-m}{l-m}}{\binom{\nr}{m'}}\sum_{q=1}^{\qr}\smix_{q}\eta_q^l - (\frac{1}{\nr}\sum_{m=0}^{\nr} \sum_{q=1}^{\qr}\rho_{q}\eta_q^m)^2 \\ 
        & + \frac{1}{\nr^2\nc^2}\nc(\nc-1) \sum_{m, m'=0}^{\nr} \sum_{l = \max(m,m')}^{\min(m+m', \nr)}\frac{\binom{m}{l-m'}\binom{\nr-m}{l-m}}{\binom{\nr}{m'}}\sum_{q, q'=1}^{\qr}\smix_{q}\smix_{q'}\eta_{q}^{l-m'}\eta_{q'}^{l-m}\eta_{qq'}^{m+m'-l}\,.
\end{align*}
\end{enumerate}

\end{proposition}

The proof is provided in the supplementary material \ref{S:proof:var}. 
The expectations of the robustness under the biSBM given in Proposition \ref{prop:rob_unif} and \ref{prop:rob_block} only rely  on the model parameters $\param$ and the number of rows $\nr$, whereas the expression of the variance also involves the number of columns $\nc$. 
Note that for the block sequences $\Sbb = \Blbb$, the calculus is not so simple, but the value could still be estimated by simulations if required.

$\var_{\inc} (\roba{\inc}{\Ubb})$ quantifies the variability of the robustness statistic among a population of networks distributed as biSBM. However, from an ecological point of view, it could also be interesting to quantify the variability of the robustness of a network with respect not only to the network but also to the extinction sequence. This is obtained by computing   
$$\var_{\inc,S}\left(\frac{1}{\nr} \sum_{m=1}^{\nr}\robasm{\inc}{S}{m}\right) =     \var_{\inc,S}\left(\robas{\inc}{S}\right) $$  

 \begin{remark}
 This total variance can be reformulated as: 
\begin{eqnarray}
 \var_{\inc,S}\left(\robas{\inc}{S}\right)  &=& \E_{S}(\var_{A|S}\left(\robas{\inc}{S}\right)) +  \var_{S}(\E_{\inc|S}(\robas{\inc}{S}\,\,)) = \E_{S}(\var_{A}\left(\robas{\inc}{S}\right))  \label{eq:var_total1} \\
 &=&\E_{\inc}(\var_{S}\left(\robas{\inc}{S}\right)) +  \var_{\inc}(\E_{S}(\robas{\inc}{S}\,\,)) \label{eq:var_total2}
 \end{eqnarray}
because $S$ and $\inc$ are independent under $\mathcal{L}_{\btheta,\bn,\Ubb}$ and $\E_{\inc|S}(\robas{\inc}{S})$ does not depend on $S$ by exchangeability. 
The total variance can be expressed explicitly for $\Sbb = \Ubb$ as 
\small{
\begin{align*}
\var_{\inc,S}&\left(\robas{\inc}{S}\right)  =  \frac{\nc}{\nr^2\nc^2}\sum_{m, m'=0}^{\nr}\sum_{q = 1}^{\qr} \smix_q \eta_q^{\nr - \min(m, m')} - \biggl(\frac{1}{\nr}\sum_{q=1}^{\qc} \smix_{q} \frac{\eta_q - \eta_q^{\nr+1}}{\con_{+q}} \biggr)^2 \\+ 
&\frac{\nc(\nc -1)}{\nr^2\nc^2}\sum_{m, m'=1}^{\nr}\sum_{q, q'=1}^{\qr}\smix_q \smix_{q'} \eta_q^{\max(m, m')-\min(m, m')}\eta_{qq'}^{\nr - \max(m, m')} \,,
\end{align*}}
through the computation of Equation \eqref{eq:var_total1}.
The second term in Equation \eqref{eq:var_total2} is provided in Proposition \ref{prop:variance}. Thus we are able to compute the remaining terms to understand the various sources of variability (due to $S$ or $A$). 
\end{remark}

\subsection{Illustration of the variability of the robustness function}

\begin{figure}[ht]
    \centering
    \includegraphics[scale = .6]{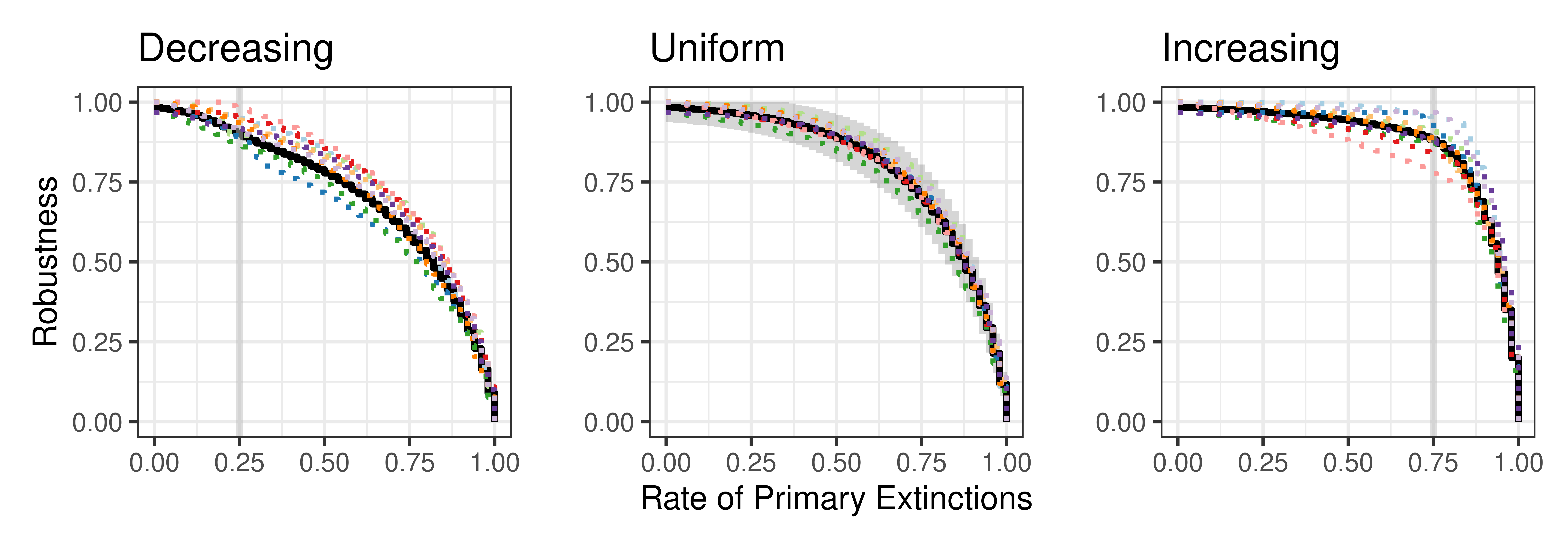}
    \caption{Robustness function computed from a set of biSBM  parameters (plain black) and by Monte Carlo for $10$ networks generated from the same biSBM distribution (dotted) for block decreasing, uniform and block increasing primary extinction sequences. The grey ribbon on the uniform facet is twice the standard error given in Proposition \ref{prop:variance}.}
    \label{fig:rob_function}
\end{figure}

In Figure \ref{fig:rob_function}, we illustrate the variations of the robustness function for a given structure encoded in 
$$\btheta= \left(\delta=\left(\begin{array}{cc}
.4 &.15\\                                                                                                                                                
.25 &.05                                                                                                                                                   \end{array}
 \right),
 \pi = (0.25,0.75) ,\rho=(0.2,0.8)
 \right)\,.$$
This corresponds to a so-called core-periphery structure where the first blocks (in rows and columns) are well connected (the core) and the second blocks are less connected (the periphery). 
We represent the functions $m \mapsto \robm{\lbm{\param}{\bn}}{\mathbb{S}}{m}$ with $\mathbb{S}=\mathbb{U}$ and $\mathbb{S}=\mathbb{B}$  such that the blocks are ordered by decreasing or increasing connectivity (solid black lines in Figure \ref{fig:rob_function}).
For the uniform distribution, we also plot the area given by $ m \mapsto \robm{\lbm{\param}{\bn}}{\mathbb{U}}{m} \pm 2 \sqrt{\var_{\inc}(\robam{\inc}{\Ubb}{m})}$ (grey ribbon). The colored dotted lines are  Monte Carlo estimates  of the robustness functions $\widehat{\rob}_{\Sbb}\left(\inc,m\right)$ corresponding to 10 simulated networks
for the same extinction sequence distributions.
The Monte Carlo estimates are computed over $300$ realizations of extinction sequences.
The inflection points and the dispersion of the dotted lines observed respectively around an extinction rate of $0.25$ for decreasing extinction and $0.75$ for increasing extinction correspond respectively to the proportion of the 
core or the periphery blocks. When the rate of extinction exceeds one of these proportions, the extinctions which were only happening in one block  then
happen in the other block.
Also notice that when $\Sbb = \Blbb$ ordered by increasing connectivity, some networks still have a robustness of $1$ even after a large fraction of primary extinctions has occurred. The robustness for this primary extinction distribution is highly dependent on the degree of the most connected primary species.

\section{Impact of the Network Structure on the Robustness}\label{sec:properties}

From the expressions derived in the previous section when $  A \sim biSBM(\btheta,\bn)$, we are now able to 
study the average behavior of the robustness with respect to the mesoscale structure of the network encoded in $\btheta$.  

\subsection{Analytical Properties}

\subsubsection{First properties on   $R_{\mathbf{\theta},\mathbf{n},\mathbb{S}}(m)$} 

We first derive the following straightforward  but  useful properties of the robustness function and statistic:

\begin{properties}{}\label{prop:rob_unif_function_lbm}       

    \begin{enumerate}
        \item Under the joint distribution $\mathcal{L}_{\param,\bn,\mathbb{S}}$ where $\Sbb \in \{\Ubb, \Blbb\}$, the following properties hold:
        \begin{enumerate}
         \item      the function
$m  \in \{0, \dots, \nr\} \mapsto \robm{\lbm{\param}{\bn}}{\Sbb}{m}$ is a strictly decreasing function provided that $\con_{k+}>0$ for all $k>0$,        
                  \item $\robm{\lbm{\param}{n}}{\Sbb}{0} \leq 1-(1-d)^{\nr} \leq  1$  where $d=  \sum_{k,q} \pi_k \rho_q\con_{kq}$,
        \item For $\param=(\pi,\rho,\con)$ and $\param'=(\pi',\rho',\con')$  such that $\pi=\pi'$, $\rho=\rho'$ and $\forall (k,q) \in \{1, \dots, Q_r\} \times \{1, \dots, Q_c\}$ $\con_{kq} \leq \con'_{kq}$, we have
       $$\forall m\in{0,\ldots,n_r}, \robm{\lbm{\param}{n}}{\Sbb}{m}\le \robm{\lbm{\param'}{n}}{\Sbb}{m} \quad \text{and}\quad \robst{\param,{\bn}}{\Sbb} \leq \robst{\param,{\bn}}{\Sbb}\,.$$
 \end{enumerate}
        \item Under the joint distribution $\mathcal{L}_{\param,\bn,\Ubb}$, if $\nr \leq \nr'$ then
         $\robst{\lbm{\param}{(\nr,\nc)}}{\Ubb} < \robst{\lbm{\param}{(\nr',\nc)}}{\Ubb}$.
    \end{enumerate}
\end{properties}

\begin{proof}
    Property  \emph{1$.$(a)} comes from the robustness definition and from the fact that $s(m+1:\nr) \subset s(m:\nr)$ for any extinction sequence $s$. Property \emph{1.(b)} is a consequence of Proposition \ref{prop:max_rob}. Property \emph{1.(c)} is true for all $m\in \{0,\ldots,n_r\}$ 
    from Equations \eqref{eq:rob_fun_unif} and \eqref{eq:robm n theta B}.
    
    For Property \emph{2.}, first notice that $ \robm{\lbm{\param}{(\nr+1,\nc)}}{\Ubb}{m+1} = \robm{\lbm{\param}{(\nr,\nc)}}{\Ubb}{m}$. Then,
    \begin{eqnarray*}
       \robst{\lbm{\param}{(\nr+1,\nc)}}{\Ubb} &= &\frac{1}{\nr+1}\Bigl(\robm{\lbm{\param}{(\nr+1,\nc)}}{\Ubb}{0} + \sum_{m = 0}^{\nr} \robm{\lbm{\param}{(\nr+1,\nc)}}{\Ubb}{m+1}\Bigr) \\
       & = & \frac{1}{\nr+1}\Bigl(\robm{\lbm{\param}{(\nr+1,\nc)}}{\Ubb}{0} + \sum_{m = 0}^{\nr} \robm{\lbm{\param}{(\nr,\nc)}}{\Ubb}{m}\Bigr) \\ 
       & = & \frac{1}{\nr+1}\Bigl(\robm{\lbm{\param}{(\nr+1,\nc)}}{\Ubb}{0} + \nr \robst{\lbm{\param}{(\nr,\nc)}}{\Ubb}\Bigr) \\
       \text{(Property  1.(a))} & > & \frac{1}{\nr+1}
       \Bigl(\robm{\lbm{\param}{(\nr,\nc)}}{\Ubb}{0} + \nr \robst{\lbm{\param}{(\nr,\nc)}}{\Ubb}\Bigr) \\
        & > & \frac{1}{\nr+1}\Bigl(\robst{\lbm{\param}{(\nr,\nc)}}{\Ubb} + \nr \robst{\lbm{\param}{(\nr,\nc)}}{\Ubb}\Bigr) = \robst{\lbm{\param}{(\nr,\nc)}}{\Ubb}
    \end{eqnarray*}

\end{proof}
Properties \emph{1.(a)} and \emph{1.(b)} supply a bound depending on the expected density and the number of row species $\nr$. 
Property \emph{1.(d)} implies that for a given block structure, the robustness increases when each element  $\con_{kq}$ increases
while Property \emph{2.} states that the uniform robustness automatically increases with the size $\nr$ of the network. 
However, it is important to note that these properties do not assess that the robustness is an increasing function of the connectance  $d$.

\subsubsection{Upper bound for the robustness under a uniform extinction sequence}

We now aim at identifying mesoscale structures maximizing the average robustness under the joint distribution $\mathcal{L}_{\param,\bn,\Ubb}$.  
In order to remove the effect of the mean number of interactions a.k.a. the density 
$d=\sum_{k,q} \pmix_k \con_{kq} \smix_{q}$, we propose to compare structures encoded in $\param$ leading to the same density value.
We also fix the number of row species $\nr$.
Thus, for a given density $d$, we  define the set $\Theta_d = \{\param=(\pi,\rho,\con) : \sum_{k,q} \pmix_k \con_{kq} \smix_{q} = d\}$.
The following proposition provides  an upper bound for the expectation of the robustness function and of the robustness statistic as a function of the density.
It also identifies a condition on $\param$ to achieve this upper bound.
Eventually, it shows that the parametrization of an Erd\H{o}s-Rényi distribution satisfies this condition and reaches  the lowest variance of the robustness statistic among the parametrizations  satisfying this condition.

\begin{proposition}{Upper bound of robustness under $\mathcal{L}_{\param,\bn,\Ubb}$.}\label{prop:max_rob}

\begin{enumerate}
  \item For all $m \in \{0 \dots, \nr\}$:
$ \robm{\param,(\nr, \nc)}{\Ubb}{m} \leq  1-(1-d)^{\nr-m}$.
 
 \item For all $m \in \{0 \dots, \nr-2\}$ 
\begin{eqnarray*}
\arg \max_{\param \in \Theta_d} \robm{\lbm{\btheta}{(\nr, \nc)}}{\Ubb}{m} & = & \left\{ \btheta : \sum_{k=1}^{\qr}\pmix_{k}\con_{kq} = \sum_{k=1}^{\qr}\pmix_{k}\con_{kq'}, \quad \forall (q, q') \in \{1, \dots, \qc\}^2 \right\}:= \Theta_{d, \nr}^{\max},
\end{eqnarray*}

Moreover, $\forall \btheta \in \Theta_d$ and $\forall \nc$: 
    \begin{equation}
       \robst{\lbm{\btheta}{(\nr,\nc)}}{\Ubb} \leq 1-\frac{1}{\nr}\frac{(1-d)-(1-d)^{\nr+1}}{d}.
    \end{equation}
  \item Assume that $E$ is a network with $n_r$ rows and $n_c$ columns following an Erd\H{o}s-Rényi distribution with density parameter $d$. Then:
  $$\min_{\inc \sim biSBM(\param,\bn):
  \param\in\Theta_{d, \nr}^{\max}}  \var_{\inc}(\roba{\inc}{\Ubb}) =  \var_{E}(\roba{E}{\Ubb})\,.$$ 
  In other words, among all parameters $\btheta \in \Theta_{d, \nr}^{\max}$, the one of the Erd\H{o}s-Rényi network minimizes the variance of robustness statistic $\roba{\inc}{\Ubb}$ given in Proposition \ref{prop:variance}.
  
\end{enumerate}
\end{proposition}

\begin{proof}

\begin{enumerate}
  \item  Recall that 
  $  \robm{\param,{(\nr, \nc)}}{\Ubb}{m}= 1- \sum_{q=1}^{\qc} \smix_q \left(1 - \con_{+q}\right)^{\nr-m}$
where $\con_{+q} = \sum_{k = 1}^{\qr}\pmix_{k}\con_{kq}$ for any $q \in \{1, \dots, \qr\}$.  
So, 
$ \robm{\lbm{\btheta}{(\nr, \nc)}}{\Ubb}{\nr} = 0$ and for $ \robm{\lbm{\btheta}{(\nr, \nc)}}{\Ubb}{\nr-1} = 1 - (1- d)$ : the bound is true for $m=\nr$ and $m = \nr-1$. Now, if  $ 0 \leq m \leq \nr -2$, then  $x 
\mapsto x^{\nr-m}$ is a strictly convex function and the Jensen. inequality applies:
\begin{eqnarray}\label{ineq:jensen}
        \robm{\lbm{\btheta}{(\nr, \nc)}}{\Ubb}{m} &=& 1- \sum_{q=1}^{\qc} \smix_q \left(1 - \con_{+q}\right)^{\nr-m} 
      \leq   1- \left( \sum_{q=1}^{\qr} \smix_q  (1 - \con_{+q})\right)^{\nr-m}  \\
      &=& 1- \left( 1 - \sum_{q=1}^{\qr} \smix_q  \delta_{+q} \right)^{\nr-m} = 1-(1-d)^{\nr-m} \nonumber,
    \end{eqnarray}

\item Equality at line \eqref{ineq:jensen} holds  if and only if the term inside the strictly convex function is constant, ie. for any $q, q'$, 
$\con_{+q} = \con_{+q'}$ .

  \item When we compute the variance given  in Proposition  \ref{prop:variance}(2)
  on any biSBM with $ \theta \in \Theta^{\max}_{d, \nr}$, we notice that
  $\eta_q = \eta_{q'} = 1 - d$ and that only the term $ \sum_{q, q'}\smix_{q}\smix_{q'}\eta_{q}^{l-m'}\eta_{q'}^{l-m} \eta_{qq'}^{m+m'-l}$ varies. 
  We can reformulate this quantity and use the Jensen inequality: 
  \begin{small}
  \begin{align*}
    \sum_{q, q'}\smix_{q}\smix_{q'}\eta_{q}^{l-m'}\eta_{q'}^{l-m}&\eta_{qq'}^{m+m'-l}  =     (1-d)^{l-m'}(1-d)^{l-m}\sum_{q, q'}\smix_{q}\smix_{q'}\eta_{qq'}^{m+m'-l} \\
    \text{(Jensen)} & \geq (1-d)^{2l-m'-m} \biggr(\sum_{q, q'}\smix_{q}\smix_{q'}\sum_{k} \pmix_k(1-\con_{kq})(1 - \con_{kq'})\biggl)^{m+m'-l} \\
    \text{(Distributivity)}& = (1-d)^{2l-m'-m} \biggr(\sum_{k} \pmix_k(1-\sum_{q}\smix_{q}\con_{kq})(1 - \sum_{q'}\smix_{q'}\con_{kq'})\biggl)^{m+m'-l} \\
    \text{(Jensen)} & \geq (1-d)^{2l-m'-m} \biggl(\Bigl(\sum_{k} \pmix_k(1-\sum_{q}\smix_{q}\con_{kq})\Bigr)^2\biggr)^{m+m'-l} \\
    & = (1-d)^{m'+ m} \qquad \text{(Initial term from ER)}
  \end{align*}
  \end{small}
\end{enumerate}

\end{proof}

\begin{figure}
  \centering
  \includegraphics[scale = .7]{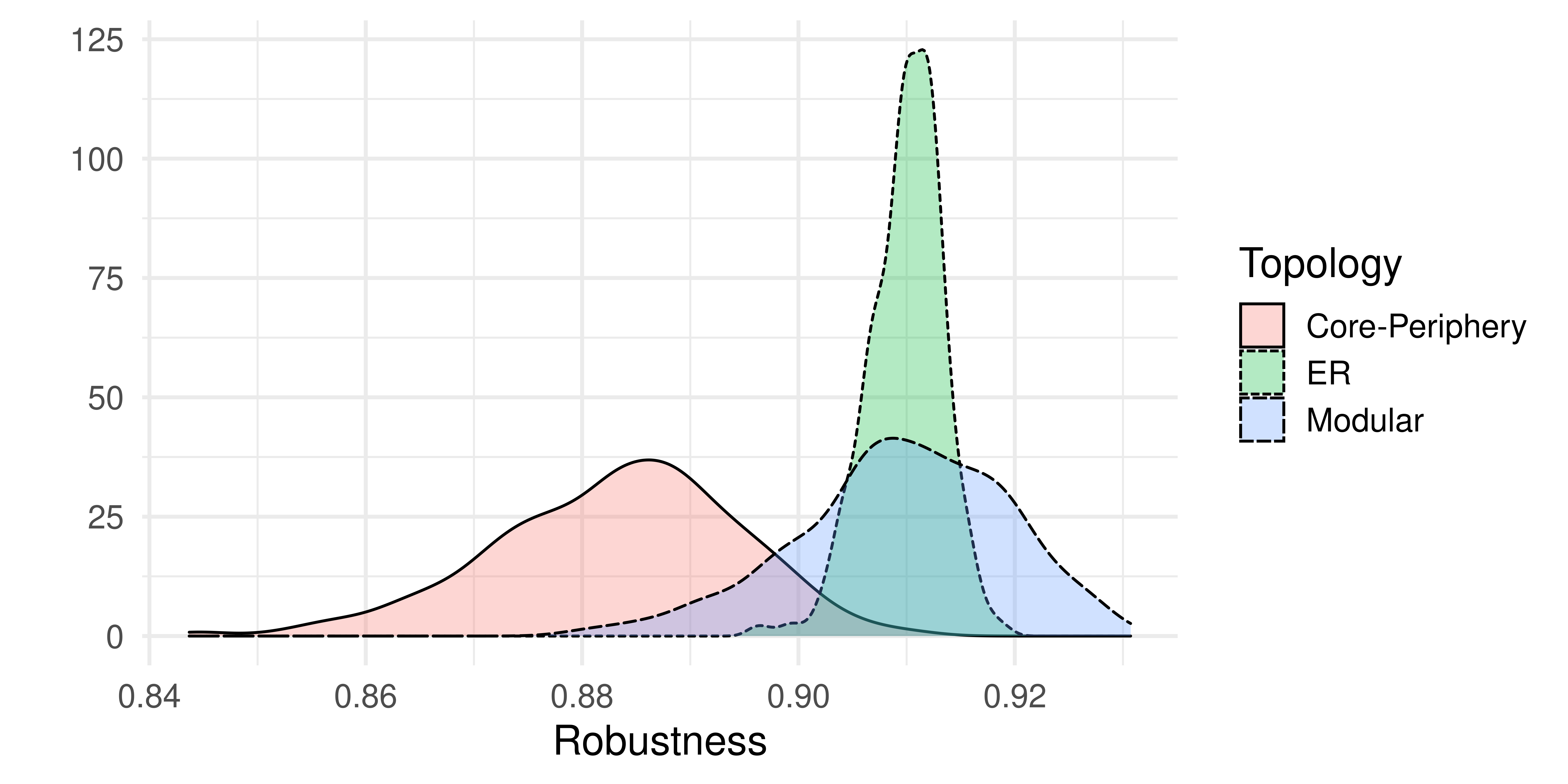}
  \caption{Density curve of the distribution of the expectation of the robustness statistic under biSBMs of the same size and density with three different sets of parameters. Each network is simulated from a given set of parameters,  then the robustness is computed by the Monte-Carlo approximation given in \eqref{eq:rob_stat2}. 500 simulations each. 
}\label{fig:variance}
\end{figure}

Although the homogeneous distribution on the networks (Erd\H{o}s-Rényi) leads to the maximum robustness in expectation for a given density, a particular realization of a network according to another distribution may be likely to have a larger robustness than a realization according to the Erd\H{o}s-Rényi distribution. Indeed, the variance is larger when it corresponds to a distribution that represents a more complex structure.
This behavior is illustrated in Figure \ref{fig:variance}.

    \subsection{Analysis for Typical Structures}

We now illustrate numerically  how the  robustness statistic varies with respect to the network topology  \footnote{All the simulations and estimations in this section are done using the \texttt{R} package \texttt{robber} \citep{robber}available on \texttt{CRAN} and documented  at \url{https://chabert-liddell.github.io/robber/}. 
 }.
For that purpose, we fix   $\nr = \nc = 100$, $\qc = \qr = 2$, and $\pmix = (.25, .75)$ and  make $\smix$ vary.  For  $j \in [1/8,8 ]$,  we consider  the following connectivity matrices.  
    \begin{description}
        \item [Modular:] 
        $\con = \begin{pmatrix} j & 1 \\
                                                    1 & j
                                    \end{pmatrix}$, \OA{}{e}ach block of row species is strongly connected to a block of column species and lowly connected to the other.
        \item [Core-periphery: ] 
        $\con = \begin{pmatrix} j & j \\
                                 j & 1 \end{pmatrix}$.
           \begin{itemize}
               \item For $j > 1$, the structure is nested, the core is strongly connected to the whole network while the periphery is lowly connected with itself. 
               \item For $j<1$, the core is strongly connected with itself but the rest of the network is lowly connected.
           \end{itemize}
    \end{description}

    Each connectivity matrix is then normalized such that the density of the network is equal to $0.0156$, by applying the following transformation:
   $
        \tilde{\con}_{kq} = 0.0156 \frac{\con_{kq}}{\sum_{k',q'} \pmix_{k'} \con_{k'q'} \smix_{q'}} \quad \forall k, q \in \{1, 2\}
   $. This value is chosen so the robustness statistic associated with an Erd\H{o}s-Rényi distribution with density $d=0.0156$ and $n_r=100$ rows is approximately equal to $0.5$. We recall this value is an upper bound of the expectation of the robustness statistic under the joint distribution $\mathcal{L}_{\btheta,\bn,\Ubb}$ where $\param$ is such that the network has the same density and same number of rows.

\begin{figure}
 \centering
    \includegraphics[scale=.6]{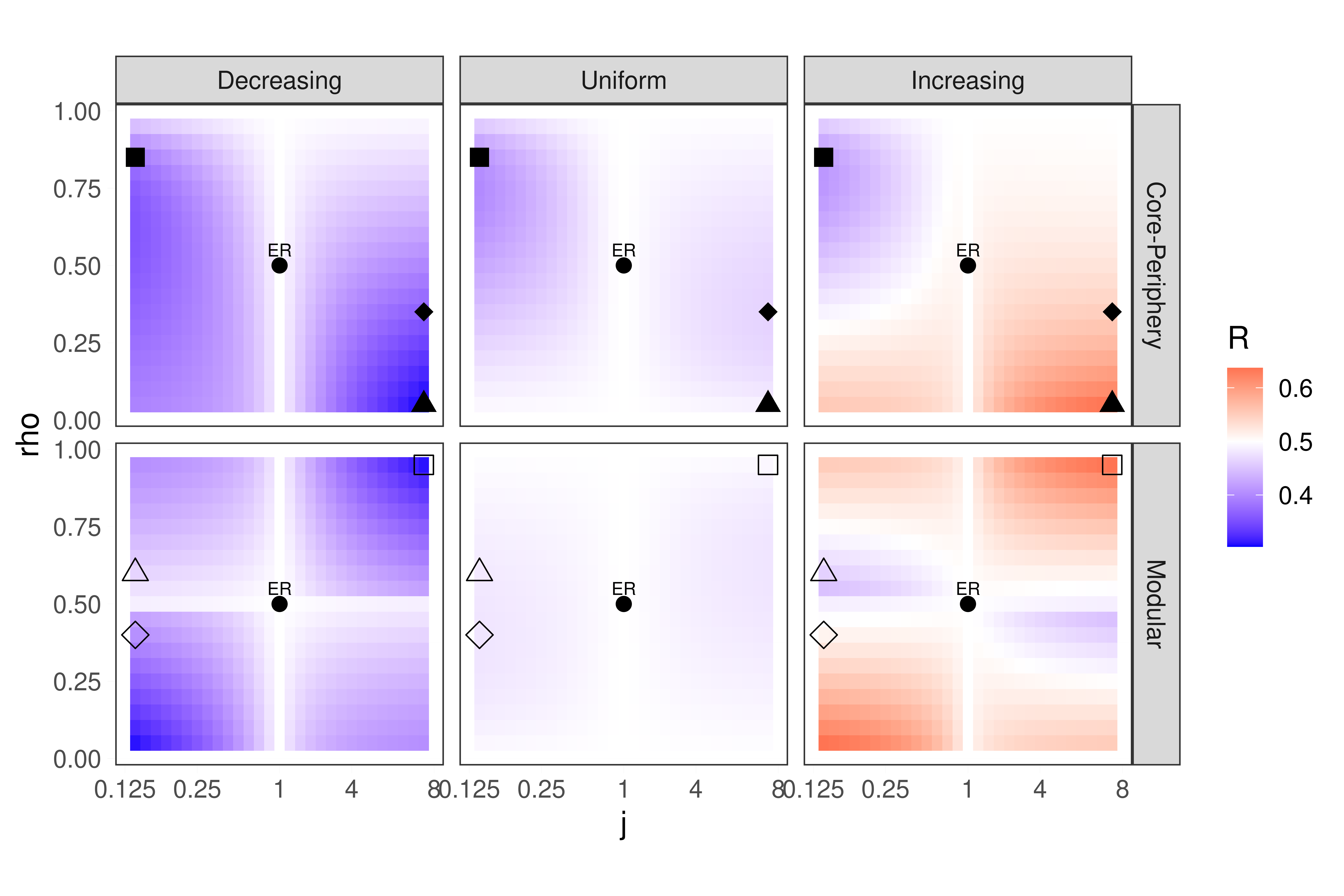}
    \caption{Robustness for different biSBM topologies.  Primary extinction sequences are displayed in rows and topologies in columns.  In abscissa,  $j$ is a topology strength parameter. \texttt{rho} is the proportion of column species that belongs to the first column block. $\bullet$ represents a topology with no structure (Erd\H{o}s-Rényi (ER)). Color gradient varies from blue (less robust than an ER), to white (as robust as an ER),  to red (more robust than an ER). The symbols corresponds to the following mesoscale structures: 
    $\blacksquare$ - large core, strongly connected only to itself, $\blacklozenge$ - small core, strongly connected to the whole network, $\blacktriangle$ - medium size core strongly connected only to itself, $\square$ - highly modular with unbalanced blocks' sizes, $\lozenge$ - highly modular with slightly smaller highly connected blocks, $\triangle$ - highly modular  with slightly larger highly connected blocks.
    }\label{fig:heatmap}
\end{figure}

 The extinction sequence distributions are  $\Sbb \in \{ \Ubb,\BbbU,\BbbD\}$ where $\BbbU$ (resp. $\BbbD$) corresponds to the block increasing (resp. decreasing) extinction sequences distribution defined in Equation \eqref{eq:Block extinc}. 
 For every $\Sbb$, topology (modular or core-periphery) and value of $j$ and $\rho$, we   compute the expectations of the robustness statistics by using the expressions derived in Section \ref{sec:moment}. 
These expectations are displayed in the heat maps of Figure \ref{fig:heatmap}.

For modular networks, the plots are symmetric with respect to the central dot which corresponds to the case of an Erd\H{o}s-Rényi distribution. The impact of the modular structure is slighter for $\Sbb  = \Ubb$ 
  than for $\Sbb \in \{\BbbU,\BbbD\}$.  
  For $\Sbb  = \Ubb$, 
  the strongest impact is observed when the modular structure is strong and the most connected blocks are slightly larger than the least connected ones (Fig. \ref{fig:heatmap}.$\lozenge$).
  Keeping the same strong modular structure with the most connected blocks slightly smaller than the least connected ones (Fig. \ref{fig:heatmap}.$\triangle$) leads to a negative impact on the robustness no matter $\Sbb$. 
  
  When the network is highly modular and the most connected block is small ($j$ and $\rho$ both small or both large, Fig. \ref{fig:heatmap}.$\square$) the robustness is strongly impacted. This impact is negative for $\BbbD$  
  and positive with $\BbbU$. 
  
  For a core-periphery structure, there is a clear asymmetry between $j<1$ and $j>1$. The robustness statistic tends to be smaller when the core is mainly connected to itself, especially when the core is small  ($j<1$ and large $\rho$: Fig. \ref{fig:heatmap}.$\blacksquare$) no matter $\Sbb$. On the other hand, core-periphery structure with small core that are highly connected to the whole network have the strongest impact on the robustness, negatively when $\Sbb = \BbbD$ 
  and positively when $\Sbb  = \BbbU$ 
  (Fig. \ref{fig:heatmap}.$\blacktriangle$). The effect of this structure ($\blacktriangle$)  is very slight on uniform extinction sequences whereas the effect  tends to get larger when the blocks' sizes are more balanced (Fig. \ref{fig:heatmap}.$\blacklozenge$).

\section{Analysis of a collection of observed bipartite ecological networks}\label{sec:appli}

In this section,  we analyse the robustness of a collection of 136 plant-pollinator, seed-dispersal or host-parasite networks,  issued from the Web of Life dataset (\url{www.web-of-life.es}).
The selected networks involve at least $10$ row species and $10$ column species. 
\\

When observing an interaction network, one can compute its empirical  robustness as defined in Equation \eqref{eq:rob_mc}. However, understanding the behavior of the robustness statistic for  any network with the same probabilistic distribution (i.e.\ with the same mesoscale structure) may also be attractive and informative. 
This can be done by positing a model on the network of  interest, estimating the corresponding parameters (summarizing its structure) and deriving the moments of the robustness for these inferred parameters using Section \ref{sec:moment} (either using a closed-form expression or  a Monte Carlo integration, depending on the model).
This expected version of the standard robustness may be considered as a new robustness indicator.   
\\

In what follows, we put in perspective  the empirical robustness and its expected versions for the biSBM and DCbiSBM models (Subsection \ref{data:compar}) and comment the systematic differences we observe.  
Then, we demonstrate the interest of the expected version    when the network is partially observed (Subsection \ref{data:missing}). Indeed,
when inferring the bipartite SBM, the observational process that generates the observed network with possibly missing data could be taken into account \citep{tabouy2019variational}. This which  allow us to compensate for observational biases in the empirical robustness.

\paragraph{Inferring the biSBM and DCbiSBM}
The parameters of these two models   can be inferred by a variational version of the Expectation-Maximization (EM) algorithm. The number of blocks is chosen according to an Integrated Classification Likelihood (ICL) criterion \citep{daudin}. This variational EM algorithm  is theoretically grounded \citep{bickel} and has proven its pratical efficiency \citep{daudin,mariadassou}. In practice for the inference of these models, we use the \texttt{blockmodels}  \texttt{R} package \citep{leger2015blockmodels}  and to handle missing observations the \texttt{GREMLINS} \texttt{R} package  \citep{bar2020block}, both available on CRAN .

\subsection{Computation of the Robustness for the Web of Life Dataset}\label{data:compar}

For each network $A$, on the one hand, we compute the empirical robustness $\widehat{\overline{\rob}}_{\Ubb}(\inc)$ using  $300$  Monte Carlo realisations.  
On the other hand, for each model (biSBM and DCbiSBM) we denote $\hat{\btheta}$ (resp. $\hat{\btheta}, \hat{\bgamma}$)  the estimated parameters and compute 
the expectations of the robustness under each model. For the biSBM, we also supply their variance and  define the ratio
$$Z_R(\inc)=\mid \widehat{\overline{\rob}}_{
\Ubb}(\inc) - \robst{\lbm{\hat{\btheta}}{n}}{\Ubb} \mid / \sqrt{\var_{\lbm{\hat{\btheta}}{n}}(\,\E_{\Ubb}[\,\robas{\inc}{S}|\inc \,]\,)}$$
which is the number of standard deviations separating the two computed robustness statistics. This quantity helps   assess the goodness of fit of the biSBM to the network $A$ with respect to the robustness statistic. 

\begin{figure}[ht]
    \centering{
    \includegraphics[scale=.5]{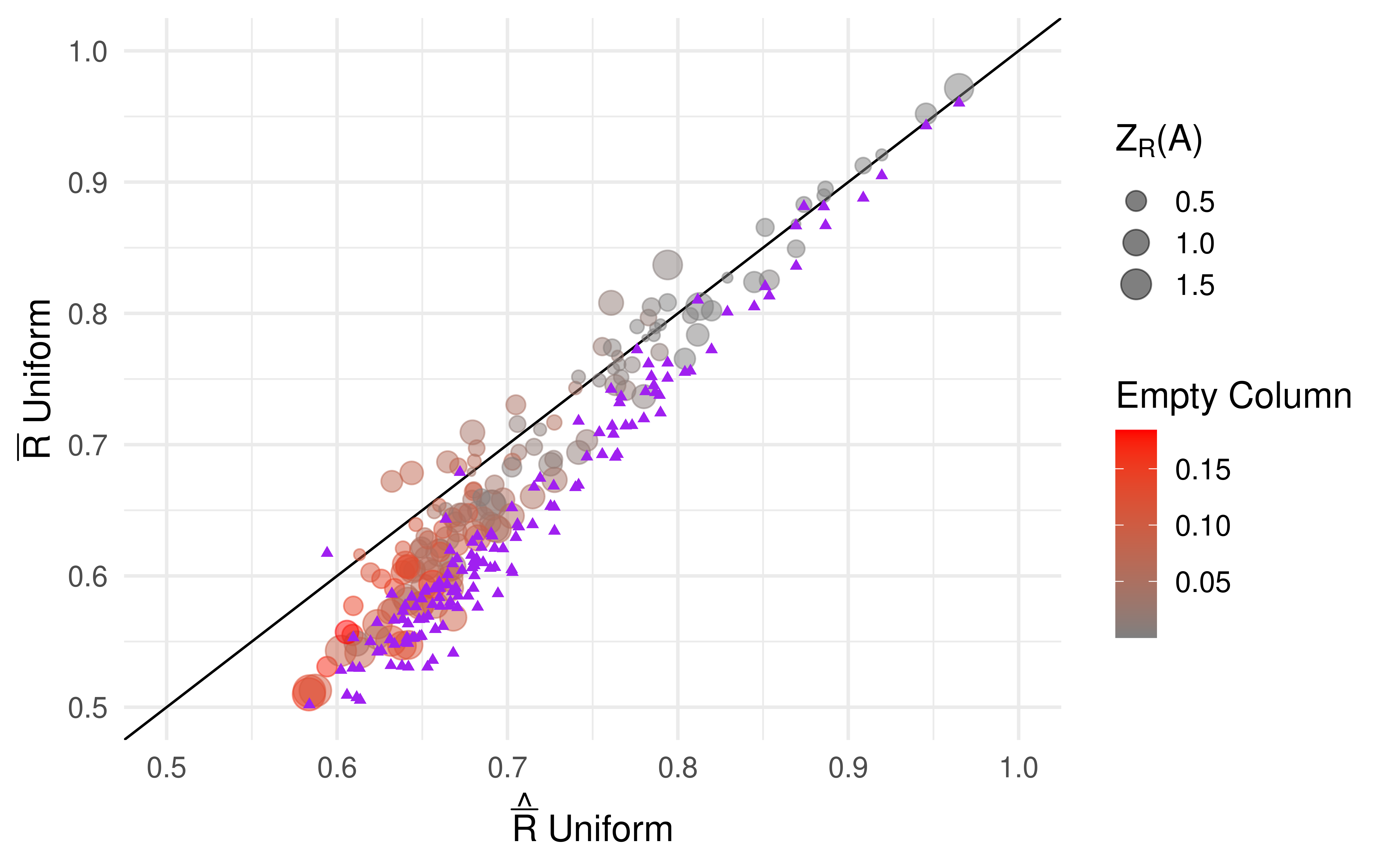}
   \caption{Expected robustness statistic ($\robst{\lbm{\hat{\btheta}}{n}}{\Ubb}$) under a biSBM (circles) and DCbiSBM (purple triangles)  as a function of the standard robustness  ($\widehat{\overline{R}}_{\Ubb}(A)$) for uniform extinction sequences. The grey to red gradient stands for $(1-\hat{d})^{\nr}$, the probability to have an empty column under an Erd\H{o}s-Rényi distribution. The size of the point is related to $Z_R(\inc)$. 
   \label{fig:gof_unif}}}
\end{figure}

\paragraph{biSBM versus DCbiSBM}
In Figure \ref{fig:gof_unif},  we plot the points $(\widehat{\overline{\rob}}_{\Ubb}(\inc), \robst{\lbm{\hat{\btheta}}{n}}{\Ubb})$  for the two models (circles are for biSBM and purple triangles for DCbiSBM).
For the biSBM, the color of the point depends on $(1-\hat{d})^{\nr}$ which is the estimated probability to observe a column with no interaction. 
The observed robustness indicators  range from $0.5$ to $1$. The points roughly follow the identity line for  the biSBM (circles) whereas the expected version seems systematically smaller than the empirical robustness for the DCbiSBM (purple triangles). We comment further \OA{}{on} these points hereafter. 

\begin{itemize}
 \item Under a biSBM, the smaller  $(1-\hat{d})^{\nr}$ , the smaller  $\mid \widehat{\overline{\rob}}_{\Ubb}(\inc) - \robst{\lbm{\hat{\btheta}}{n}}{\Ubb} \mid $ and $Z_R(A)$  (the number of standard deviations). When $(1-\hat{d})^{\nr}$ is large (red dots), the expected version of the robustness under a biSBM underestimates the standard  robustness (red dots at the bottom left of the plot).
This phenomenon may be explained as follows. 
By construction, ecological networks only involve species that have been seen at least one time in interaction (species with no interaction were removed). 
As a consequence, since the biSBM does not take into account this phenomenon, the space of networks we integrate over when computing the robustness statistic under a biSBM is too large. This remark is especially true for small networks   
and highlights the limitation of the biSBM model for small ecological networks.  
\item On $127$ networks (over the $136$), the DCbiSBM selects  $1$ block, encoding solely, as the EDD model, the degree distribution and the size of the networks in their parameters. On these networks, the expectation of the robustness statistic under a DCbiSBM  is smaller than the empirical robustness most of the time  (purple triangles). We can conclude that the DCbiSBM seems to be unable to capture the additional structure beyond their degree distribution that makes them more robust to uniform random extinction than expected.
\end{itemize}

This first study highlights the limitation of the DCbiSBM model to mimic ecological networks (hence, hereafter, we  focus our analysis solely on the biSBM). On the contrary, for not too small networks, the expected version  of the robustness with biSBM  supplies coherent robustness values with the empirical indicator $\widehat{\overline{\rob}}_{\Ubb}(\inc)$. This new version of the robustness has the advantage to arise in a closed-form and a quantification of its variance is available. This comparison has been made for $\Sbb = \Ubb$, we now consider more elaborate extinction sequence distributions.

\begin{figure}[ht]
    \centering{
    \includegraphics[scale=.5]{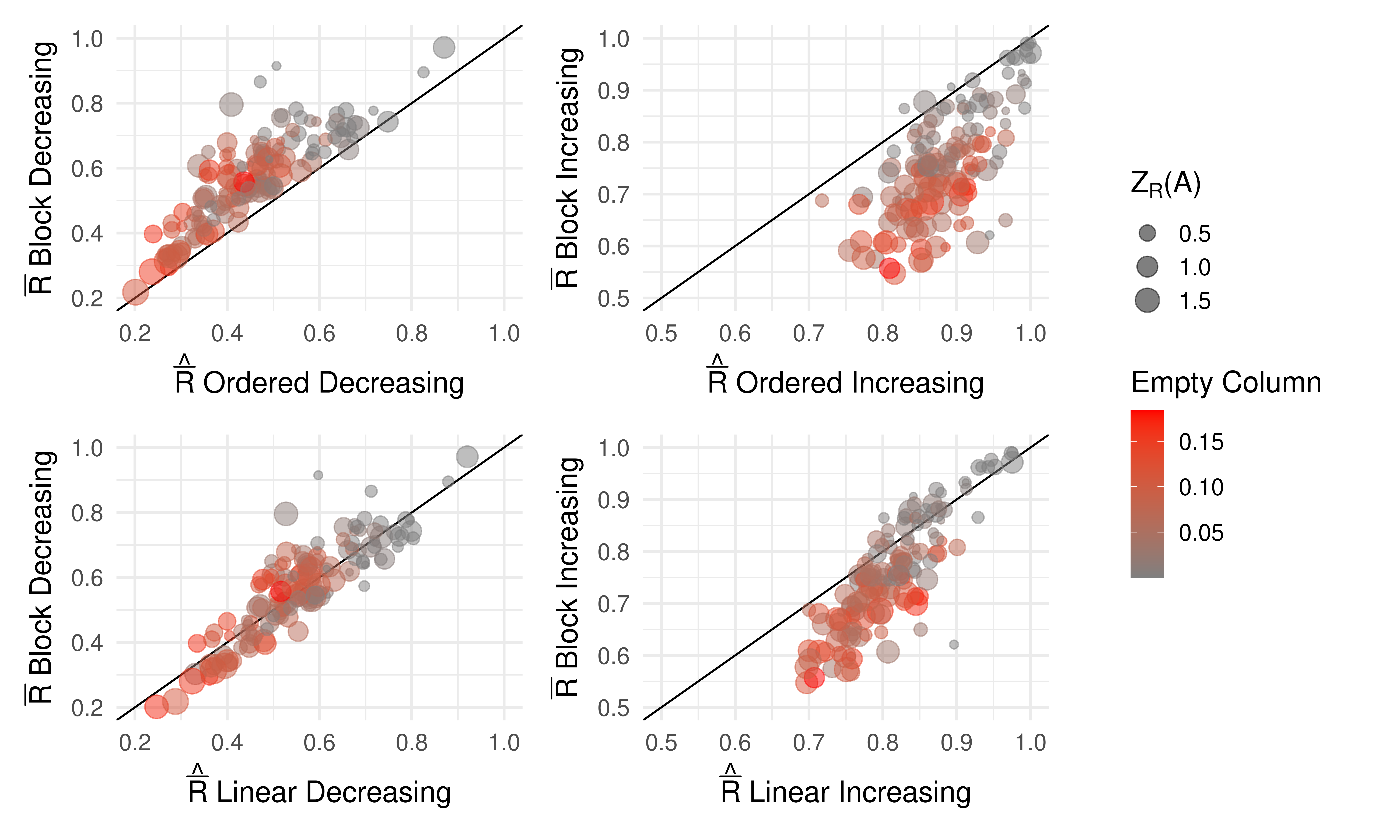}
   \caption{$\robst{\lbm{\hat{\btheta}}{n}}{\Sbb} $     as a function of the classical robustness $\widehat{\overline{R}}_{\Dbb}(\inc)$ for various $\Sbb$. \textsf{Block Decreasing} (resp. \textsf{increasing})  = $\BbbD$ (resp $\BbbU$), \textsf{Ordered Decreasing} (resp. \textsf{Increasing}) = $\DbbDo$ (resp. $\DbbUo$). 
   \textsf{Linear Decreasing} (resp. \textsf{Increasing}) = $\DbbDa$ (resp. $\DbbUa$). 
The grey to red 
gradient stands for $(1-\hat{d})^{\nr}$, the probability to have an empty column under an Erd\H{o}s-Rényi distribution. The 
size of the point is the deviation between the biSBM and the empirical robustness in terms  of the number of standard deviations of the robustness under a biSBM distribution for $\Sbb = \Ubb$.
\label{fig:gof_block}}}
\end{figure}

\paragraph{About $\Sbb$. }
We compute the robustness statistics for decreasing and increasing primary extinction sequences on the degrees with the two methods described in Section \ref{sec:robustness}: the one  which strictly depends on the order of the degrees ($\widehat{\overline{R}}_{\DbbUo}(\inc)$ and $\widehat{\overline{R}}_{\DbbDo}(\inc)$) and  the one where the nodes are weighted by a linear function of the degrees as in Equation \eqref{eq:weight_degree} with $\alpha = 1$ ($\widehat{\overline{R}}_{\DbbUa}(\inc)$ and $\widehat{\overline{R}}_{\DbbDa}(\inc)$).  
We examine the fit with  $\robst{\lbm{\hat{\btheta}}{n}}{\Sbb} $  where $\Sbb \in \{\BbbD, \BbbU \}$ mimics the degree increasing and decreasing extinction sequences. The comparison are  plotted in Figure \ref{fig:gof_block}.

When considering primary extinction sequences by decreasing connectivity, there is a positive bias of $\robst{\lbm{\hat{\btheta}}{n}}{\BbbD}$ on  $\widehat{\overline{R}}_{\DbbDo}(\inc)$ (top-left) which is attenuated when considering  $\widehat{\overline{R}}_{\DbbDa}(\inc)$ (bottom-left). For primary extinction sequences by increasing connectivity, the empirical robustness is highly dependent on the degrees of the most connected species, hence the fit of $\robst{\lbm{\hat{\btheta}}{n}}{\BbbU}$ on $\widehat{\overline{R}}_{\DbbUo}(\inc)$ is very poor with a negative bias (top-right). While the negative bias still remains, the fit  on $\widehat{\overline{R}}_{\DbbUa}(\inc)$ is correct for a much higher  fraction of the networks (bottom-right).\\

As a conclusion, the degree decreasing sequences standardly used in the ecological field can be easily reproduced with our block decreasing connectivity sequences, leading to quite comparable values. Once again, our expected version of the robustness can be calculated in a closed-form while the empirical  robustness relies on a computationally expensive Monte Carlo integration.  The degree decreasing extinction sequences are very sensible to the most highly connected species leading to no agreement between  the two versions of the robustness.  

\subsection{Correction for Partially Observed Networks}\label{data:missing}

Although the ecological networks are often considered to describe all the possible interaction between species, the sampling may be incomplete
\citep{bluthgen2008interaction} which may bias the computed network statistics \citep{rivera2012effects} such as the robustness.
By relying on a probabilistic model such as the biSBM which can be adjusted to account for the observation process, the sampling effect on the robustness can be corrected. 
We assume that we could have obtained the \textit{true} interaction network if the sampling effort has been large enough. Instead of this true network, we have only a partially observed interaction network that corresponds to a subset of the true network. 
We assume that one of the two following frameworks may have generated missing data:

\begin{description}
    
    \item [Partially observed species]  $25\%$ of both the row species and the column species are removed uniformly at random, resulting in the following networks subset:
    $A^{obs} = \{\inc_{ij} : i \text{ and } j \text{ are observed}  \}$.
    In this framework, we need to assume that some other data or some expert knowledge gives  us the true number of species and the density of the true network. By relying on this expert knowledge, we are able to adjust the parameters of the biSBM.

    \item [Partially observed interactions]
  The observation process consists of the observation of interaction on two different transects $T_1$ and $T_2$. Since not all the species are present on both transects, the interactions between the species which were not observed on the same transect are labelled as missing and encoded by $\texttt{NA}$s. 
  This results in the following modified incidence matrix: 
  
   $A^{obs} =  \begin{bmatrix} & j \in \{T_1 \cap T_2\}  & j \in \{T_1 \setminus T_2\} & j \in \{T_2 \setminus T_1\} \\
                       i \in \{T_1 \cap T_2\}     &  A_{ij} & A_{ij} & A_{ij} \\
                       i \in \{T_1 \setminus T_2\}& A_{ij}& A_{ij}& \textcolor{red}{\texttt{NA}}\\
                       i \in \{T_2 \setminus T_1\}& A_{ij}& \textcolor{red}{\texttt{NA}} & A_{ij}\end{bmatrix}$.

    More precisely, we consider that on average $50\%$ of the species were observed on both transects while $25\%$ were just observed on one of the two transects and select those species uniformly at random, resulting in $12.5\%$ of missing interactions on average.
  Note that in this case, we do not need any expert or additional knowledge to have an unbiased estimation of the biSBM parameters. It is sufficient that the missing values are taken into account in the biSBM inference.
  
\end{description}

\begin{figure}[!ht]
    \centering{
    \includegraphics[scale=.6]{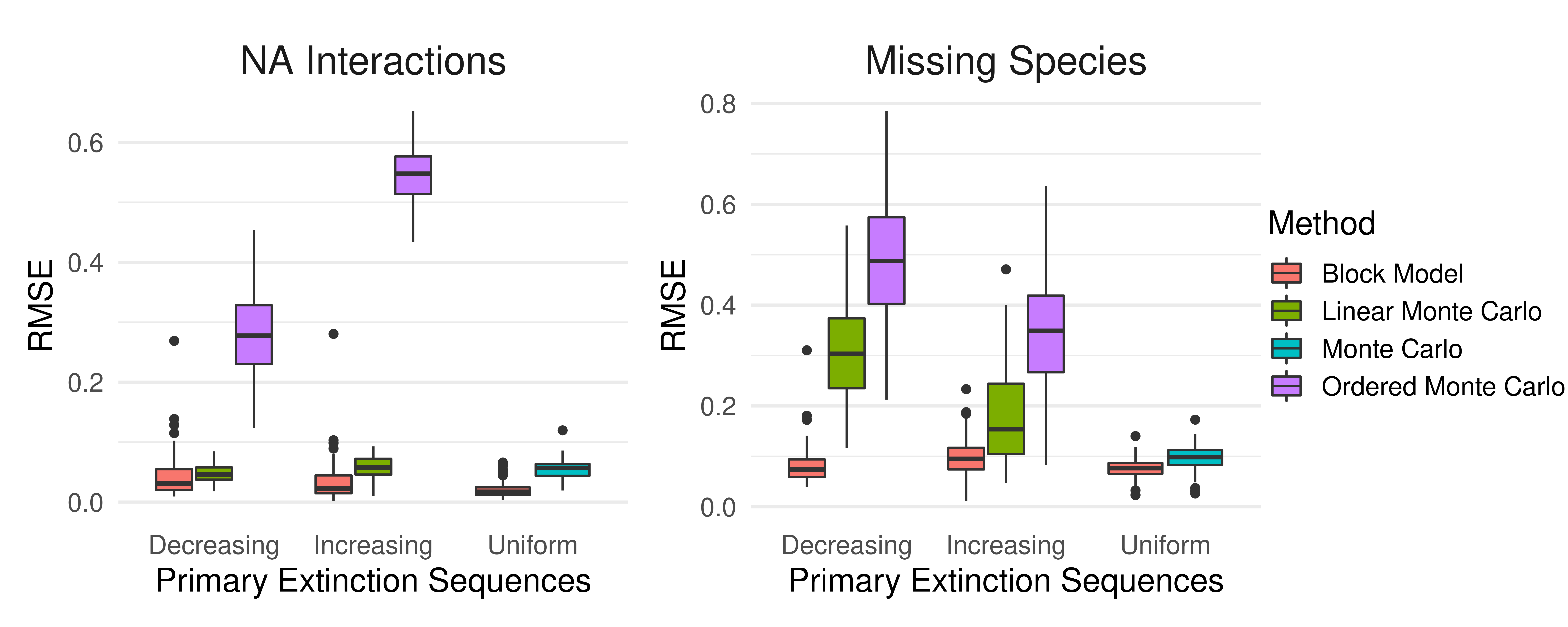}
    \caption{Error (RMSE) in the prediction of the robustness of $98$ fully observed networks computed from partially observed networks.
    On the left, $12.5\%$ of possible interactions are recoded as \texttt{NA}.
    and on the right,  $25\%$ of species are missing.  \label{fig:wol_pred}}}
\end{figure}

We perform\OA{}{ed} a simulation study where we consider\OA{}{ed}
  98 networks  from the web of life dataset as the \textit{true} interaction networks.
  We \OA{}{kept} the 98 networks where $(1-d)^{\nr} < .1$ and the error in terms of standard deviation is smaller than $1$. 
For each of these $98$ networks, we simulate\OA{}{d} the two observation processes described above $30$ times, then we compute\OA{}{d} the standard robustness statistic   and its expectation 
under a biSBM.
For Figure \ref{fig:wol_pred}, we compute\OA{}{d} the root mean squared error (RMSE) for each network and for different cases (in terms of observation process and extinction sequence) between the
robustness computations on the true network (fully observed) and on the partially observed network. 
When the method for computing the robustness is based on MC computations,
we consider\OA{}{ed} the uniform primary extinction sequences (named Monte Carlo in Fig. \ref{fig:wol_pred}) and the increasing and decreasing extinction sequences 
that depend strictly (named Ordered Monte Carlo in Fig. \ref{fig:wol_pred}) or proportionally as in Equation \eqref{eq:weight_degree} (named Linear Monte Carlo in Fig. \ref{fig:wol_pred}) on the row degrees. More specifically, the RMSE for the methods based on MC computations are computed for any extinction sequences $\Sbb = \{\Ubb, \DbbUo, \DbbUa, \DbbDo, \DbbDa\}$ listed above, as:
$$\sqrt{\frac{1}{30}\sum_{b=1}^{30}(\widehat{\bar{\rob}}_{\Sbb}(\inc)-\widehat{\bar{\rob}}_{\Sbb}(\tilde\inc_b))^2}$$
for each of the $98$ fully observed networks $\inc$ and where the $\tilde\inc_b$'s are realizations of a  partial observation of $\inc$.
In order to get  the robustness under a biSBM, the biSBM \OA{}{was} first inferred by taking into account missing data and the robustness statistics \OA{}{were} then computed under the inferred biSBM. 
We compute\OA{}{d} the corresponding RMSE for the distributions on extinction sequences $\Sbb \in \{ \Ubb,\BbbU,\BbbD\}$ as: 
$$\sqrt{\frac{1}{30}\sum_{b=1}^{30}(\robst{\lbm{\hat\param}{\bn}}{\Sbb}-\robst{\lbm{\tilde\param_b}{\bn}}{\Sbb})^2}$$
where $\hat\param$ is estimated from the fully observed network and 
the $\tilde\param_b$'s are estimated from partial observations of the network.
Note that the numbers of species in row and in column are the same 
for the fully and the partially observed networks even in the case of missing species since we assume\OA{}{d} that we \OA{}{had} some additional knowledge which provide\OA{}{d} us with this information.

The errors in the prediction of the robustness statistics computed from 
partially observed networks
are much smaller when computing the robustness under a biSBM than when using 
MC computations.
Indeed, the missing data can be accounted for in the biSBM inference 
and the underlying structure of the network can still be recovered from partial information whereas the Monte Carlo simulations are more sensitive to perturbations in the network.
The extinction sequences which depend strictly on the degrees are the most impacted by a partial observation of the network. This impact is rather strong  although only 
$12.5\%$ of the information is missing. Note that  adding some randomness in the primary extinction sequences by using a distribution which depends linearly on the degrees instead of a strict order has a stabilizing effect. 

    \OA{}{
    In the framework with partially observed species we assumed that we had additional knowledge giving us access to the true number of species and network density.  In practice, this is not always the case.  When details about the sampling of networks are available, a few methods exist in the literature to estimate the number of species \citep{jimenez2006comparing, gotelli2011estimating} and similar methods could be used to estimate the density. One could then plug these estimates in the formula of robustness.  If neither the sampling scheme nor the expert knowledge is available, the true number of species remains unknown. Note that in most available ecological interaction networks, the number of species is underestimated as species which have not been seen in interaction with other species are not included in the network. Finally, having access to the number of species, but not to the true number of interactions, leads to an underestimated robustness.
    }

\section{Discussion}

\OA{}{W}e proposed an expected version of the empirical robustness for  bipartite ecological interaction networks by considering a joint model on the network and on the primary extinction sequences. In particular, we obtained \OA{}{a} closed and tractable form of the robustness when  considering a biSBM as the network model for uniform and by blocks primary extinction sequences. We validated our method by showing that the obtained values were consistent with the empirical robustness classically used in the ecological network community. Having analytical forms allows us to better understand the impact on the robustness of the topology of the network in terms of  number of species, density\OA{}{,} and mesoscale connectivity patterns.
Furthermore, we used the difference between the empirical and expected versions 
to show that the biSBM is better suited than the DCbiSBM for bipartite ecological networks. 
This could come from the fact that the DCbiSBM tends to favor community structure above core-periphery structure \citep[14.7.3]{newman2018networks}. The core-periphery structure is indeed very common in bipartite ecological networks which has a strong impact on the robustness.

On real networks, this method can serve as an alternative to the traditional empirical approach, especially when networks are partially observed (because of the inference stability of the biSBM) or when the sampling effort is incomplete (because the model parameters can be easily corrected). This step could be improved  with more precise information by considering specific sampling schemes in the model \citep{tabouy2019variational} or by obtaining more details on the sampling in order to better estimate the parameters block by block.  Moreover, from the observation of a network, the impact of some hypotheses on the structure might be tested by tuning some parameters of the models and computing the corresponding expected robustness.
It may also help study the impact of the structure beyond 
the effect of the number of species or density when comparing the robustness of several networks: indeed, once the biSBM parameters are estimated, the robustness statistics can be computed by setting the same numbers of species and the same density for all the networks.

Although we developed our method for bipartite networks, this approach can be extended to other types of networks in particular multipartite networks \citep{pocock2012robustness} and food webs. 
 Multilayer networks (multiplex or multipartite networks) are gaining a lot of attention with scholars studying ecological networks \citep{hutchinson2019seeing}. Extending this framework to multipartite networks by including cascading effect between layers is an interesting perspective once data will be more readily available.
 The study of robustness for food webs is quite active but  requires some ecological insight to properly model the food web as some species are basal species and thus \OA{}{do} not prey on other species (in-degree is 0). So food webs which are usually blockmodeled by a directed stochastic block model, might be better handled in our case using a multipartite block model with basal species as a functional group of its own. In this case, it might be important to incorporate rewiring or cascading mechanisms to  the modeling of the extinctions. \OA{}{A direction to look at in order to deal with the extinction of basal species and the incorporation of cascading mechanisms is the approach of Bayesian networks to model species extinction in food webs \citep{eklof2013secondary,haussler2020bayesian}.}
 
Lastly, we believe that other ecological indicators could be estimated through a parametric model based approach. Especially, the EDD model and the DCbiSBM seem particularly well suited to study nestedness \citep[see][for a review]{mariani2019nestedness} which is a widely used statistic for ecological networks.

\section*{Aknowledgement}
The authors would like to thank Sonia Kefi, François Massol and Vincent Miele for their helpful advice. 
This work was supported by a public grant as part of the Investissement d'avenir project, reference ANR-11-LABX-0056-LMH, LabEx LMH. This work was partially supported by the grant ANR-18-CE02-0010-01 of the French National Research Agency ANR (project EcoNet).

\bibliographystyle{chicago}
\bibliography{biblio}

\appendix
\renewcommand{\thesection}{S-\arabic{section}}

\section{Proof for Section 4 (Moments of the robustness statistic)}

\setcounter{proposition}{1}

\begin{proposition}
Let $(\inc,S) \sim \mathcal{L}_{\btheta,\bn,\Blbb}$, then
\begin{equation}\stepcounter{equation}\tag{S-{\theequation}}\label{S:eq:robm n theta B}
  \robm{\lbm{\param}{\bn}}{\Blbb}{m} = 
    1 - \sum_{q=1}^{\qc} \smix_q \sum_{n_1 + \dots + n_{\qr} = \nr} \frac{\nr!}{n_1! \dots n_{\qr}!}  
\prod_{k=1}^{\qr}\pmix_{k}^{n_k} \left( 1-\con_{kq}\right)^{\min^{+}(n_k , \underset{l \leq k}{\sum}n_l - m)},
\end{equation}
where $\min^{+}$ is the positive part of the minimum function:  $\min^{+}(x, y) = \max(0, \min(x,y))$. 
\end{proposition}

\begin{proof}\label{S:proof:block}
  Let us assume without loss of generality that the primary extinction sequences go from block $1$ to block $\qr$. Under this assumption,
  $
    \pr(S = s | \pcl) = \frac{\mathbf{1}_{\pcl_{s(1)} \leq \dots \leq \pcl_{s(\nr)}}}{\# \{s : \pcl_{s(1)} \leq \dots \leq \pcl_{s(\nr)}\}}
  $.
Thus,  an extinction sequence which does not maintain the block ordering has a null probability. Then, conditioning first by the blocks memberships then the primary extinction sequences:
\begin{small}
  \begin{align}\stepcounter{equation}\tag{S-{\theequation}}\label{eq:rob_by_block_proof}
    \E_{(\inc,S)}&[\robasm{\inc}{S}{m}]  = \sum_{q_{1:\nc}}^{\qc}\smix_{1:\nc}\sum_{k_{1:\nr}}^{\qr}\pmix_{1:\nr}\E_{{(\inc,S)}}[\robasm{\inc}{S}{m}| \pcl_{1:\nr} = k_{1:\nr}, \scl_{1:\nc} = q_{1:\nc}]\\ \nonumber
    & = 1-\frac{1}{\nc}\sum_{j=1}^{\nc} \sum_{q}^{\qc}\smix_{q}\sum_{k_{1:\nr}}^{\qr}\pmix_{1:\nr}\E_{(\inc,S)} \left[\prod_{i = m+1}^{\nr} (1-\inc_{S(i)j}) | \pcl_{1:\nr} = k_{1:\nr}, \scl_{j} = q\right]\\ \nonumber
    & = 1-\frac{1}{\nc}\sum_{j=1}^{\nc} \sum_{q}^{\qc}\smix_{q}\sum_{k_{1:\nr}}^{\qr}\pmix_{1:\nr} \sum_{s \in \Symnr}\pr(S = s \mid \pcl_{1:\nr} = k_{1:\nr})\left( \prod_{i=m+1}^{\nr} (1-\con_{k_{s(i)}q})\right).
\end{align}
  \end{small}
We have for all sequences $s, s'$ in the support of $S \mid \pcl$, that for all $i$, $\pcl_{s(i)} = \pcl_{s'(i)}$. Hence, using the exchangeability properties of the biSBM for species belonging to the same block:
$
  \prod_{i=m+1}^{\nr} \left(1-\con_{k_{s(i)}q}\right) =   \prod_{i=m+1}^{\nr} \left(1-\con_{k_{s'(i)}q}\right)
$.
Furthermore,
\begin{align*}
  \prod_{i=m+1}^{\nr} \left(1-\con_{k_{s(i)}q}\right)&  =   \prod_{i=m+1}^{\nr} \prod_{k=1}^{\qr}\left(1-\con_{kq}\right)^{\mathbf{1}_{\{k_{s(i)} = k\}}} =  \prod_{k=1}^{\qr}\left(1-\con_{kq}\right)^{\sum_{i = m+1}^{\nr}\mathbf{1}_{\{k_{s(i)} = k\}}}, 
\end{align*}
with for all $s$ such that $k_{s(1)} \leq \dots \leq k_{s(\nr)}$:
\begin{small}
  \begin{equation*}
   \sum_{i = m+1}^{\nr}\mathbf{1}_{\{k_{s(i)} = k\}} = 
    \begin{cases}
      0 & \text{if} \quad \underset{l \leq k}{\sum} n_l \leq m, \\
      \underset{l \leq k}{\sum} n_l - m & \text{if} \quad \underset{l \leq k}{\sum} n_l - n_k < m < \underset{l \leq k}{\sum} n_l,\\
      n_k & \text{if} \quad m \leq \underset{l \leq k}{\sum} n_l  - n_k.
    \end{cases}
  \end{equation*}
\end{small}
This leads to:
\begin{small}
\begin{align}
   \sum_{s \in \Symnr}\pr(S = s \mid & \pcl_{1:\nr} = k_{1:\nr})  \prod_{i=m+1}^{\nr} (1-\con_{k_{s(i)}q}) = \sum_{s \in \Symnr}\pr(S = s \mid \pcl_{1:\nr} = k_{1:\nr}) 
   \prod_{k=1}^{\qr}\left(1-\con_{kq}\right)^{\min^{+}(n_k, \sum_{l\leq k} n_l - m)} \nonumber \\
   & =    \prod_{k=1}^{\qr}\left(1-\con_{kq}\right)^{\min^{+}(n_k, \sum_{l\leq k}n_l - m)}. \stepcounter{equation}\tag{S-{\theequation}}\label{eq:rob_by_block_proof_psz}
\end{align}
\end{small}

Inputting Equation \eqref{eq:rob_by_block_proof_psz} in Equation \eqref{eq:rob_by_block_proof}, we obtain the result:
  \begin{align*}
    \E_{(\inc,S)}[\robasm{\inc}{S}{m}] & = 
    1-\sum_{q=1}^{\qc} \smix_q \sum_{n_1 + \dots + n_{\qr} = \nr} \frac{\nr!}{n_1!\dots n_{\qr}!}  \times \prod_{k=1}^{\qr}\pmix_{k}^{n_k} \left( 1-\con_{kq}\right)^{\min^{+}(n_k, \sum_{l\leq k}n_l - m)}.
  \end{align*}

\end{proof}

\begin{proposition}\label{S:prop:variance}
 Let $\inc \sim biSBM(\btheta,\bn)$,  $\eta_q = 1-\con_{+q}$ and $\eta_{qq'} = \sum_{k=1}^{\qr} \pmix_k(1-\con_{kq})(1 - \con_{kq'})$.
 
 \begin{enumerate}
   \item Then:
  \begin{eqnarray*}
    \var_{\inc}[\robam{\inc}{\Ubb}{m}] &=&\frac{1}{\nc}   \sum_{l = m}^{\min(2m, \nr)}\frac{\binom{m}{l-m}\binom{\nr-m}{l-m}}{\binom{\nr}{m}}\sum_{q=1}^{\qr}\smix_{q}\eta_q^l - (\sum_{q=1}^{\qr}\rho_{q}\eta_q^m)^2 \\ 
        && + \frac{(\nc-1)}{\nc}  \sum_{l = m}^{\max(2m, \nr)}\frac{\binom{m}{l-m}\binom{\nr-m}{l-m}}{\binom{\nr}{m}}\sum_{q, q'=1}^{\qr}\smix_{q}\smix_{q'}(\eta_{q}\eta_{q'})^{l-m}\eta_{qq'}^{2m-l}\,.
\end{eqnarray*}
   \item The variance of the robustness statistic due to the network variability under a given biSBM is:
\begin{align*}
    \var_{\inc}&[\roba{\inc}{\Ubb}] = \frac{1}{\nr^2\nc^2} \nc \sum_{m, m'=0}^{\nr} \sum_{l = \max(m,m')}^{\min(m+m',\nr)}\frac{\binom{m}{l-m'}\binom{\nr-m}{l-m}}{\binom{\nr}{m'}}\sum_{q=1}^{\qr}\smix_{q}\eta_q^l - (\frac{1}{\nr}\sum_{m=0}^{\nr} \sum_{q=1}^{\qr}\rho_{q}\eta_q^m)^2 \\ 
        & + \frac{1}{\nr^2\nc^2}\nc(\nc-1) \sum_{m, m'=0}^{\nr} \sum_{l = \max(m,m')}^{\min(m+m', \nr)}\frac{\binom{m}{l-m'}\binom{\nr-m}{l-m}}{\binom{\nr}{m'}}\sum_{q, q'=1}^{\qr}\smix_{q}\smix_{q'}\eta_{q}^{l-m'}\eta_{q'}^{l-m}\eta_{qq'}^{m+m'-l}\,.
\end{align*}
\end{enumerate}

\end{proposition}

\begin{proof} \label{S:proof:var}
    The calculus of the following proof relies on the following lemma:
\begin{cbnlem}\label{lem:cbn}
  Let $m, m'$ the first terms of two permutations of $\mathfrak{S}_n$, the the proportion of couple of permutation $(s, s')$  that have exactly $l$ unique terms is:
  \small{\begin{equation*}
    \frac{\# \{(s, s') \colon s(1:m) \cup s'(1:m')= l\}}{\#(\mathfrak{S}_n, \mathfrak{S}_n)} = \begin{cases}
                                      
\frac{\binom{m}{m+m'-l}\binom{n-m}{l-m}}{\binom{n}{m'}} & if 
\max\{m,m'\} \leq 
l \leq \min\{m+m', n\} \\
                                      0 & otherwise
                                     \end{cases},
  \end{equation*}}
  where $s(1:m) = \{s(1), \dots, s(m)\}$.
\end{cbnlem}
\begin{proof}
 There are $n!$ permutations of size n. For the first permutation of size $n$, we look at the first $m$ . In order to create the second permutation we must among the first $m'$ terms take:
 \begin{itemize}
   \item $l-m$ terms that are not common with the first permutation (among $n-m$)
   \item $m+m'-l$ terms that are common with the first permutation (among $m$).
 \end{itemize}

Those $m'$ terms can be reordered into $m'!$ possible arrangements and the $n-m'$ resting terms into $n-m'!$ permutations, giving:
\begin{equation*}
  \# \left\{(s, s') \colon s(1:m) \cup s'(1:m')= l\right\} = n!\binom{m}{m+m'-l}\binom{n-m}{l-m}m'!(n-m')!
\end{equation*}
We then divide by the $(n!)^2$ couple of permutations possible.
\begin{equation*}
    \frac{\# \{(s, s') \colon s(1:m) \cup s'(1:m')= l\}}{\#(\mathfrak{S}_n, \mathfrak{S}_n)} =  \frac{\binom{m}{m+m'-l}\binom{n-m}{l-m}}{\binom{n}{m'}}
\end{equation*}

The bound on $l$ is straightforward.
\end{proof}

We first prove the result for the robustness statistic. The variance of the  robustness function is a straightforward derivation from it. 
 \begin{small}   
    \begin{align} \stepcounter{equation}\tag{S-{\theequation}}\label{eq:var_A_E_S}
            \var_{\inc}&(\,\E_{S}[\,\robas{\inc}{S})| \inc \,]\,) =  \underbrace{\E_{\inc}[\,\E_{S}^{2}[\, 1 - \robas{\inc}{S}| \inc\,]\,]}_{B} - \underbrace{\E_{\inc}^{2}[\,\E_{S}[\, 1 - \robas{ \inc }{S}|  \inc \,]\,]}_{C} 
    \end{align}
 \end{small}
    Using Equation \eqref{eq:rob_fun_unif} and Equation \eqref{eq:rob_stat2}, we have:
 $
        C = \left(\, \frac{1}{\nr} \sum_{m=0}^{\nr} \sum_{q=1}^{\qc}\smix_q\eta_q^{\nr-m} \,\right)^2 \nonumber
$.
    We divide $B$ into $2$ terms,  based on the column index:
 \begin{small}    \begin{align}
        B & = \E_{\inc}\left[\, (\,1 - \frac{1}{\nr}\sum_{m=0}^{\nr-1} \E_{S}[\, \robasm{\inc}{S}{m}|\inc] \,)^2 \,\right] \nonumber\\ 
        & = \frac{1}{\nr^2}\sum_{m, m'=0}^{\nr-1}\E_{\inc}\left[\, 1 -  \E_{S}[\, \robasm{\inc}{S}{m}|\inc] \cdot \E_{S}[\, 1 - \robasm{\inc}{S}{m'}|\inc] \,\right] \nonumber\\
         & = \frac{1}{\nr^2}\sum_{m, m'=0}^{\nr-1}\E_{\inc}  \left[\,  \left(\,\frac{1}{\nr!} \underset{s \in \mathfrak{S}_{\nr}}{\sum} \frac{1}{\nc} \sum_{j = 1}^{\nc}\mathbf{1}_{\{\overset{\nr}{\underset{i = m+1}{\sum}} \inc_{s(i)j}=0 \}}\,\right) \right. \nonumber\\
         &  \qquad \qquad \cdot \qquad \left. \left(\,\frac{1}{\nr!} \sum_{s' \in \mathfrak{S}_{\nr}} \frac{1}{\nc} \sum_{j' = 1}^{\nc}\mathbf{1}_{\{\overset{\nr}{\underset{i = m'+1}{\sum}}  \inc_{s'(i)j'}=0 \}} \,\right) \,\right]\nonumber\\
         & = \frac{1}{\nr^2}\sum_{m, m'=0}^{\nr-1} \frac{1}{\nc^2} \sum_{j \neq j '= 1}^{\nc}\underbrace{\frac{1}{\nr! \nr!} \sum_{s, s' \in \mathfrak{S}_{\nr}} \E_{\inc}\left[\,   \mathbf{1}_{\{\overset{\nr}{\underset{i = m+1}{\sum}}  \inc_{s(i)j}=0 \}}  \mathbf{1}_{\{\overset{\nr}{\underset{i = m'+1}{\sum}}  \inc_{s'(i)j'}=0 \}} \,\right]}_{B1} \nonumber\\
         &\qquad + \frac{1}{\nr^2}\sum_{m, m'=0}^{\nr-1} \frac{1}{\nc^2} \sum_{j = 1}^{\nc}  \underbrace{\frac{1}{\nr! \nr!} \sum_{s, s' \in \mathfrak{S}_{\nr}}\E_{\inc}\left[\,  \mathbf{1}_{\{\overset{\nr}{\underset{i = m+1}{\sum}}  \inc_{s(i)j}=0 \}}  \mathbf{1}_{\{\overset{\nr}{\underset{i = m'+1}{\sum}}  \inc_{s'(i)j}=0 \}}   \,\right]}_{B2}.  \stepcounter{equation}\tag{S-{\theequation}}\label{eq:B}
    \end{align}
 \end{small}
    To compute $B1$, we separate the set of the union of the extinction sequences into $3$ disjoint sets in order to distribute the mixture parameter $\pmix$:
 \begin{small}
    \begin{eqnarray*}
        B1 & = & \frac{1}{\nr!\nr!} \sum_{s, s' \in \mathfrak{S}_{\nr}} \E_{\pmix, \smix} \left[\,\E_{\con}\biggr[\,   \mathbf{1}_{\{\overset{\nr}{\underset{i = m+1}{\sum}}  \inc_{s(i)j}=0 \}}  \mathbf{1}_{\{\overset{\nr}{\underset{i = m'+1}{\sum}} \inc_{s'(i)j'}=0 \}} | \pcl, \scl\,\biggr]\,\right] \\ 
        & = & \frac{1}{\nr!\nr!} \sum_{s, s' \in \mathfrak{S}_{\nr}} \E_{\pmix, \smix} \left[\, \prod_{i =m+1}^{\nr}( 1- \con_{\pcl_{s(i)}\scl_j} )  \prod_{i = m'+1}^{\nr} (1-\con_{\pcl_{s'(i)}\scl_{j'}}) \,\right] \\ 
        & = & \frac{1}{\nr!\nr!} \sum_{s, s' \in \mathfrak{S}_{\nr}} \sum_{q, q'} \smix_q \smix_{q'} \sum_{k_{1:\nr}}\pmix_{k_{1:\nr}} \prod_{i =m+1}^{\nr} (1 - \con_{k_{s(i)}q}) 
        \prod_{i = m'+1}^{\nr} (1 - \con_{k_{s'(i)}q'}) \\
        & = & \frac{1}{\nr!\nr!} \sum_{s, s' \in \mathfrak{S}_{\nr}} \sum_{q, q'} \smix_q \smix_{q'} \sum_{k_{1:\nr}}\pmix_{k_{1:\nr}} \prod_{i \in s(m+1:\nr)\setminus s'(m'+1:\nr)} (1 - \con_{k_{i}q})\\
        && \qquad \prod_{i \in s'(m'+1:\nr)\setminus s(m+1:\nr)} (1 - \con_{k_{i}q'}) \prod_{i \in s(m+1:\nr)\cap s'(m'+1:\nr)} (1 - \con_{k_{i}q})(1 - \con_{k_{i}q'})\\
        \text(Lem. \ref{lem:cbn}) & = & \sum_{l} \frac{\binom{m}{m+m'-l}\binom{n-m}{l-m}}{\binom{n}{m'}} \sum_{q, q'} \smix_q \smix_{q'} \eta_{q}^{l-(\nr-m')} \eta_{q'}^{l-(n-m)}\eta_{qq'}^{2n-m-m' - l}
    \end{eqnarray*}
 \end{small}
    For $B2$, the column indices are the same, hence some entries of $\inc$ are taken twice:
 \begin{small}
    \begin{eqnarray*}
        B2 &=& \frac{1}{\nr!\nr!} \sum_{s, s' \in \mathfrak{S}_{\nr}} \sum_{q} \smix_q  \sum_{k_{1:n}}\pmix_{k_{1:n}} \E_{\con}\biggr[\,   \mathbf{1}_{\{ \underset{i \in s(m+1:\nr)}{\sum} \inc_{ij}=0 \}}  \\
        &&  \qquad \cdot \mathbf{1}_{\{ \underset{i' \in s'(m'+1:\nr)}{\sum} \inc_{i'j}=0 \}} | \pcl_{1:\nr} = k_{1:\nr}, \scl_{j} = q\,\biggr] \\
        &=& \frac{1}{\nr!\nr!} \sum_{s, s' \in \mathfrak{S}_{\nr}} \sum_{q} \smix_q  \sum_{k_{1:n}}\pmix_{k_{1:n}} \\
        && \qquad \cdot \E_{\con}\biggr[\,   \mathbf{1}_{\{\underset{i \in s(1:\nr-m)\cup s'(1:\nr-m')}{\sum} \inc_{ij}=0 \}}| \pcl_{1:\nr} = k_{1:\nr}, \scl_{j} = q\,\biggr]\\
        &=& \frac{1}{\nr!\nr!} \sum_{s, s' \in \mathfrak{S}_{\nr}} \sum_{q} \smix_q  \sum_{k_{1:n}}\pmix_{k_{1:n}} \prod_{i \in s(1:\nr-m)\cup s'(1:\nr-m')} (1-\con_{k_iq})\\
        \text{(Lem. \ref{lem:cbn})} & = & \sum_{l} \frac{\binom{m}{m+m'-l}\binom{n-m}{l-m}}{\binom{n}{m'}}\sum_{q} \smix_{q} \eta_{q}^{l}
    \end{eqnarray*}
 \end{small}
    Finally we use the  symmetry between $m$ and $\nr-m$ and input first $B1$ and $B2$ into Equation \eqref{eq:B} then $B$ and $C$ into Equation \eqref{eq:var_A_E_S} to obtain the variance of the  robustness statistic.
    
    To obtain the variance of the robustness function, we fix $m$, and set $m=m'$ in Lemma \ref{lem:cbn}, and Equation \eqref{eq:B}.

\end{proof}

\end{document}